# Integration of Omics Data and Systems Biology Modeling: Effect of Cyclosporine A on the Nrf2 Pathway in Human Renal Kidneys Cells

Jérémy Hamon[1*], Paul Jennings[2], Frédéric Y. Bois[1,3]

[1] Chair of Mathematical Modeling for Systems Toxicology, Université de Technologie de Compiègne, BP 20529, 60205 Compiègne Cedex, France. (jeremy.hamon@utc.fr, frederic.bois@utc.fr)

[2] Division of Physiology, Department of Physiology and Medical Physics, Innsbruck Medical University, Innsbruck 6020, Austria. (paul.jennings@i-med.ac.at )

[3] INERIS, DRC/VIVA/METO, Parc ALATA, BP 2, 60550 Verneuil en Halatte, France.

## Abstract

*Background*: Integrating omics approaches and mathematical systems biology models is a challenge. In a recent paper, Wilmes *et al.* demonstrated a qualitative integration of omics data streams to gain a mechanistic understanding of cyclosporine A toxicity. One of the major conclusion of Wilmes *et al.* is that cyclosporine A strongly activates the nuclear factor (erythroid-derived 2)-like 2 pathway (Nrf2) in renal proximal tubular epithelial cells exposed *in vitro*. We pursue here the analysis of those data with a quantitative integration of omics data with a differential equation model of the Nrf2 pathway. That was done in two steps: (i) Modeling the *in vitro* pharmacokinetics of cyclosporine A (exchange between cells, culture medium and vial walls) with a minimal distribution model. (ii) Modeling the time course of omics markers in response to cyclosporine A exposure at the cell level with a coupled PK-systems biology model. Posterior statistical distributions of the parameter values were obtained by Markov chain Monte Carlo sampling. *Results*: Data are well simulated, and the known *in vitro* toxic effect EC50 is well matched by model predictions. *In vitro* pharmacokinetic results bring out the presence of nonlinear phenomena in the distribution of cyclosporine A in the cell. At the same time, systems biology results bring out the presence of an important dose-dependent response. *Conclusions*: Our modeling and simulations of the CsA mediated ROS production gives biologists insight into mechanisms of toxicity and provide quantitative estimates of toxicity beyond the time and dose range used in experiments.

## Abbreviations

**AhR**: aryl hydrocarbon receptor (transcription factor protein)
**ARE**: antioxidant response element
**ARNT**: aryl hydrocarbon receptor nuclear translocator (protein)
**CsA**: cyclosporine A
**CsA$_{cytosol}$**: Cystosol CsA quantity
**CsA$_{extracellular}$**: Extracellular CsA quantity
**CsA$_{wall}$**: CsA on wall quantity
**CYP**: cytochrome P450 3A5
**DRE**: dioxin response element
**GCL**: glutamate cysteine ligase
**GCLC**: glutamate cysteine ligase catalytic subunit
**GCLM**: glutamate cysteine ligase modifier subunit
**GPx**: glutathione peroxidase
**GS**: glutathione synthetase
**GSH**: glutathione
**GSSG**: glutathione disulfide
**Keap1**: Kelch-like ECH-associated protein 1
**MCMC**: Markov chain Monte Carlo
**mRNA**: messenger ribonucleic acid
**MRP**: Multidrug resistance protein
**MTX**: metabolomic
**Nrf2**: nuclear factor (erythroid-derived 2)-like 2
**NMA**: complex of Nrf2-Maf-ARE
**NRS**: Non reactive species
**PK**: pharmacokinetics
**PTX**: proteomic
**ROS**: reactive oxygen species
**RPTEC**: renal proximal tubular epithelial cells
**TCX**: transcriptomic
**XAA**: complex of CsA-AhR-ARNT
**XAAD**: complex of CsA-AhR-ARNT-DRE
**zmol**: zeptomole (1 zmol = $10^{-21}$ mol)
**γ-GC**: γ-glutamylcysteine

## Introduction

The quantitative modeling of toxicity pathways is a topic of current interest in predictive pharmacology and toxicology [2,3,4]. One of its challenges is to integrate omics data with systemic biology models for parametric inference and model checking [5]. In a recent paper, Wilmes *et al.* [1] demonstrated a qualitative integration of transcriptomic (TCX), proteomic (PTX) and metabolomic (MTX) data streams to gain a mechanistic understanding of cyclosporine A (CsA) toxicity. CsA is an important molecule for immunosuppressive treatment and is used in many post-graft medical protocols [6]. However, at high dose it is nephrotoxic, causing damage to the kidney vasculature, glomerulus and proximal tubule [7,8,9]. Yet, the precise mechanisms of its toxicity are still unclear: CsA is thought to induce oxidative stress at the mitochondrial level, and co-administration of antioxidants with CsA appears to mitigate its nephrotoxic effects [10].

The Nrf2 oxidative response pathway is triggered when oxidative stress is sensed by Keap-1, resulting in stabilization and nuclear translocation of Nrf2 [11]. Nrf2 binds to the antioxidant response element (ARE) inducing the transcription of several genes involved in glutathione synthesis and recycling, antioxidant activity and phase II metabolism and transport [11]. The Nrf2 response has been shown to induced in several tissues in response to chemical and physiological stressors. The kidney and particularly the proximal tubule is especially sensitive to oxidative stress and many nephrotoxins induce Nrf2 nuclear translocation and Nrf2 dependent gene induction in renal epithelial cells, including potassium bromate, cadmium chloride, diquat dibromide and cyclosporine A [1,12,13]. Moreover, we have recently shown physiological stress such as glucose depletion and subsequent re-introduction results in Nrf2 activation in renal cells [14]. Here, we pursue here the analysis of that data set with a quantitative integration of omics data with a systems biology model of the Nrf2 pathway. The model predictive ability is then assessed.

**Methods**

*Data*

RPTECs culture conditions, CsA concentrations measurements, and TCX, PTX and MTX data collection and analysis were described in detail in Wilmes *et al.* [1]. Briefly, RPTECs cells were cultured in serum-free medium and matured for two weeks on microporous supports. They were then treated for fourteen days with daily doses of CsA. The assay medium was renewed prior to each dosing. Three groups of assays were performed in triplicate: control, low CsA concentration (5 μM) and high CsA concentration (15 μM).

CsA concentration was measured in the medium on the first day at 0.5h, 1h, 3h, 6h and 24h (just before changing the medium), on the third, fifth, seventh, and tenth day at 24h (before changing the medium), and on day fourteenth at the same times than on the day one. Intracellular (cell lysate) CsA concentration and quantity bound to plastic were measured on the first and last days at the same times. Samples for TCX (conducted with Illumina® HT 12 v3 BeadChip arrays), PTX (conducted with HPLC-MS) and MTX (conducted with direct infusion MS) measurements were obtained at the end of day 1, day 3 and day 14. All fold-changes were calculated using the absolute value measured at the first time of the control experiment as a reference, and for all doses. Typical RPTEC cell volume was determined by electron microscopy and stereology, to be 2000 ± 140 μm$^3$.

All the data used are given in Supplementary tables S1 to S8.

## *Mathematical model*

Modeling was done into two steps: (i) Modeling the i*n vitro* pharmacokinetics (PK) of CsA (exchange between cells, medium and vial walls) with a minimal distribution model. (ii) Modeling the effects of CsA on omics markers as the cellular level with a coupled PK-systems biology model.

In vitro *pharmacokinetic model.* A 3-compartment model was developed to describe CsA exchange between cell medium, cells and vial walls [1]. In that model, CsA can enter and exit the cells, bind to and unbind from the plastic walls and can be metabolized within cells. Several mathematical forms for exchange rates were tested. The best fit was obtained using a first order entry into cells with Michaelis-Menten (saturable) exit rate, a first order attachment to vial wall with non-integer (fractal) order detachment, and Michaelis-Menten metabolism. The following differential equations were used to describe the time course of CsA quantities in the cytosol, medium, and on vial walls:

$$\frac{\partial CsA_{cytosol}}{\partial t} = CL_{in_1} \frac{CsA_{extracellular}}{V_{extracellular}} - \frac{CL_{out_1} \cdot CsA_{cytosol}}{V_{cytosol} \cdot Km_{out_1} + CsA_{cytosol}} - \frac{v_{max} \cdot CsA_{cytosol}}{Km_2 + CsA_{cytosol}} \quad (1)$$

$$\frac{\partial CsA_{extracellular}}{\partial t} = -CL_{in_1}\frac{CsA_{extracellular}}{V_{extracellular}} + \frac{CL_{out_1} \cdot CsA_{cytosol}}{V_{cytosol} \cdot Km_{out_1} + CsA_{cytosol}}$$
$$- k_1 \cdot CsA_{extracellular} + k_2 (CsA_{wall})^{k_3} \quad (2)$$

$$\frac{\partial CsA_{wall}}{\partial t} = k_1 \cdot CsA_{extracellular} - k_2 (CsA_{wall})^{k_3} \quad (3)$$

The model parameters are described in Table 1.

*Coupled PK-systems biology model of the Nrf2 pathways* (Figure 1). The model used was adapted from Zhang *et al.* [15]. The model full set of equations is provided as supplemental material. In brief, CsA induces oxidative stress by increasing reactive oxygen species (ROS) production. ROS, owing to their electrophilicity, can be detected by the molecular sensor Kelch-like ECH-associated protein 1 (Keap1), which promotes the ubiquitination and eventual degradation of Nrf2 [16,17]. When Keap1 is oxidized, Nrf2 ubiquitination is lowered [17], making Nrf2 available to enter the nucleus. Once in the nucleus, Nrf2 binds to small Maf proteins to form Nrf2-Maf heterodimers [18]. Those can bind to antioxidant responses elements (ARE) in the promoter region of glutamate cysteine ligase catalytic subunit (GCLC), glutamate cysteine ligase modifier subunit (GCLM), glutathione synthetase (GS), glutathione peroxidase (GPx), and MRP genes, inducing their transcriptions [16,18]. GCLC, GCLM, and GS are involved in GSH synthesis. GPx detoxifies ROS, using GSH as a co-substrate. Zhang's model was developed for a generic xenobiotic, so the following structural changes were made to consistently describe the cell kinetics and mode of action of CsA:

- CsA can enter or exit the cell, and attach to or detach from the vial walls as in the *in vitro* PK model (eqs. A10, A11, A13). Inside the cell, CsA distribution to the nucleus is explicitly modeled (eq. A12);
- In the cell, CsA is metabolized by cytochrome P450 3A5 (CYP3A5) into a metabolite X′ (not followed and non-influent on the system). CsA is mainly metabolized by CYP3A isoforms [6], and in kidney cells only CYP3A5 is significantly expressed [19];
- Oxidative stress, characterized by the total quantity of oxidative compounds in the cell (ROS), was explicitly introduced in the model as a state variable (eq. A75);

- The production of ROS depends on CsA concentration in the cell, and ROS are eliminated by GPx (eq. A75) in a non-reactive species pool (NRS) (not followed and non-influent on the system);
- Keap1 and the Nrf2-Keap1 complex are oxidized by ROS (eqs. A52, A53, A72, A73).

All the other equations are the same than in Zhang *et al.* [15]. In addition, some parameters had to be set to particular values for CsA: In the model, xenobiotics can bind to the AhR nuclear receptor. However, CsA is not a known AhR ligand, so its binding parameters ($k_{b_2}$ and $k_{b_5}$) were set to zero. Supplemental Table S9 gives the model parameters and the state variables initial values.

### *Calibration of the models*

The *in vitro* PK model parameters were calibrated, using Bayesian inference [5,20], with data on the CsA quantities measured in the medium, cells, and on vial walls (Table S1 to S6). Non-informative (vague) prior parameter distributions were used (see Table 1). The data likelihoods were assumed to follow a lognormal distribution around the model predictions, a standard assumption with such measurements. Measurement error geometric standard deviations were assumed to be specific to each of the three measurement types (different procedure were used for their obtention). They were assigned vague log-uniform prior distributions and were calibrated together with the other model parameters. A total of 11 parameters (8 kinetic parameters and 3 statistical ones) were calibrated. Posterior statistical distributions of the parameter values were obtained by Markov chain Monte Carlo (MCMC) sampling [5].

For the coupled PK-Nrf2 pathway model, the parameters directly controlling CsA kinetics were set to the joint posterior distribution mode found by the above calibration (see Table 1). Another set of 27 structural parameters (Table 2) was calibrated using fold-change omics data (as a function of time and CsA dose) on Nrf2 mRNA, CYP3A5 mRNA, GS mRNA, GCLC mRNA, GCLM mRNA, GST mRNA, GPx mRNA, MRP mRNA, GCLM protein, GS protein, MRP protein, γ-glutamylcysteine (γ-GC), and GSH. Four of those parameters have a direct influence on the rate of ROS synthesis, metabolism and interaction with Keap1. Another 15 parameters controle the activation and induction of Nrf2, GCLC, GCLM, GST, GPx, CYP3A5, GS and MRP genes transcription. Another six parameters controle synthesis and elimination of γ-GC and GSH, and two last parameters controle CsA metabolism and Nrf2 and Maf binding. Model predicted fold-changes were computed the same way as the experimental ones, using the actual quantity predicted at the first time of the control experiment as a reference. The prior parameter distributions chosen were either vague or centered around the values used by Zhang [15] (see Table 2). The data likelihoods were assumed to be lognormal distributions around the model predictions. The same measurement error geometric standard deviation was assumed for all omics measurements. It was calibrated together with the other model parameters, using a vague log-uniform prior. Here also, posterior distributions of the parameter values were obtained by MCMC sampling. For each model parameter sampled, convergence was evaluated by computing the potential scale reduction criterion of Gelman and Rubin [21] on the last 200,000 iterations from each simulated chain.

### *Software used*

All model simulations and MCMC calibrations were performed with GNU MCSim v5.4.0 [5]. The R software, version 2.15.1 [22] was used for other statistical analyses and plots.

## Results

Results for the *in vitro* pharmacokinetic model have been previously reported in Wilmes *et al.* [1] and are briefly summarized here. Overall, the data were well simulated. Exposure to low concentrations of CsA (5 μM) led to a dynamic steady state in about 3 days. The average ratio between extracellular and intracellular CsA concentrations was about 200, consistent with the fact that CsA is lipophilic and accumulates in cells. Exposure to high concentrations of CsA (15 μM) led to major alterations of the biodistribution of CsA over time. Steady state was reached in the cells only after approximately 7 days. The average ratio of intracellular to extracellular CsA concentrations was about 200 on the first day (like at low concentration) and around 650 on the last day. Those results highlight the presence of nonlinear phenomena in the distribution of CsA in RPTECs. Note here the importance of administering repeated doses: A unique administration would not have uncovered those phenomena.

Figure 2 show the influence of the extra-cellular concentration of CsA on the evolution of intra-cellular CsA quantities over time. Those results confirm the presence of nonlinear phenomena in the distribution of CsA in the cell. Moreover, above about 15 μM extra-cellular CsA, intra-cellular concentration do not reach a plateau within 14 days. This can be explained by the saturation of the cell efflux mechanism.

*Coupled PK-systems biology model of the Nrf2 pathway.* All the analyses and predictions presented here were made using a (posterior) random sample of 5000 parameter vectors, obtained by thinning the 200,000 last iterations of five MCMC chains at convergence. Figures 3 and 4 show the model fit obtained for the *in vitro* omics data, at low and high CsA exposure dose, respectively. The bundle of curves presented reflect residual uncertainty in the model predictions, resulting from unavoidable measurement errors and modeling approximations. Overall the data is reasonably well simulated. For all species, the time profiles are clearly different between the two doses. At low CsA exposure (Figure 3) periodic oscillations are pervasive and after 14 days, the system does not appear to have reached a dynamic equilibrium. For many curves (corresponding to parameter vectors, including the most probable one), the oscillations are not stable. Their period may differ from one curve to another and go up to four days, even though the period of CsA administration is exactly one day. Supplemental Figure S1 extends the simulation period to 60 days, time at which a dynamic equilibrium is reached in all cases, and shows the same oscillation pattern. At the high CsA exposure (Figure 4) two patterns emerge:

The first type of profile, which concerns all species except GSH and γ-GC, is a plateauing curve. Different maximum values are reached after three days by different curves. The second profile type, which only concerns GSH and γ-GC, begins by oscillations which have not stabilized within 14 days. Supplemental Figure S2 extends the simulation period to 60 days and precises the behavior of GSH and γ-GC. The initial oscillations decrease gradually in the amplitude to completely disappear after about 30 days.

Figure 5 illustrates the evolution of free nuclear Nrf2 protein and cellular ROS quantities over 14 days at low or high repeated CsA dosing. These are model predictions for two non-observed species. Additional simulations were performed up to 60 days and the trends were similar (data not shown).

As for the previous species for which we had data, large differences are seen between low and high dosing. At low CsA exposure a cyclic pattern is observed, which disappears at high exposure where the ROS quantity grows (less than exponentially) while the Nrf2 protein quantity reaches systematically a plateau.Figure 6 shows the influence of the extracellular CsA concentration on the evolution of cellular ROS, nuclear Nrf2 protein, cellular GSH and cellular GCL quantities over time. Figures for other species (GCLC, GCLM, GPx, GS and GST) are not shown because their profiles are very similar to the GCL one. Figure 6 shows that the extracellular concentration of CsA has a large influence on the amount of ROS in the cytosol. For extracellular CsA concentrations below 8 μM CsA, the concentration (or quantity) *vs*. time profile of cytosolic ROS is oscillating, above 8 μM CsA, the ROS profile rises in a hockey-stick fashion.

For nuclear Nrf2, cellular GSH and cellular GCL, depending on the extracellular CsA concentration, the model predicts either oscillating profiles or plateauing concentration *vs*. time profiles. As for ROS, the transition is rather abrupt and occurs approximately at 8 μM CsA.

**Discussion**

A proper assessment of drug or chemical safety from *in vitro* assays requires the measurement of concentration of the parent molecule and eventually its metabolites) in the assay medium and in cells [23,24]. Kinetic modeling can then be used to interpolate and extrapolate the data obtained. Here, an *in vitro* pharmacokinetic model was built using LC-MS/MS data on the distribution of CsA over time in human RPTECs. CsA is highly lipophilic and its rapid uptake and accumulation in cells was observed. At 5 μM CsA (daily initial extracellular concentration), the model indicated that steady-state was reached in about 2 days, whereas at 15 μM CsA, steady-state was reached only after 7 days. Moreover, cellular CsA concentrations at steady-state were clearly not proportional to exposure, and a disproportionate accumulation of CsA was observed at high exposure. Drug accumulation in target tissues is often associated with tissue-specific toxicities, and it is important to account for it. Such a bioaccumulation could be explained by the saturation of the P-glycoprotein efflux transporter (ABCB1) above 5 μM exposure [1]. We did not observe a modulation of CsA PK by its PD in our *in vitro* system, even though CsA interactions with transporters are known [25]. In particular, CYP 3A5 levels were not affected by CsA levels, so CsA metabolism was not disturbed by induction or repression.

Zhang's model was not intended to be used specifically with CsA or our cell system, so we had to re-calibrate several parameter values. This was done in a Bayesian statistical framework [20], to take into account the prior information we had on several parameters. The choice of parameters to be estimated was based on the results of a sensitivity analysis performed on all parameters in the model (results not shown). For most parameters, the *posterior* mean estimate was clearly different from the *prior* mean. Compared to Zhang's model, the parameters controlling ROS formation ($kf_{75}$) and ROS elimination ($v_{max8b}$) stabilize at, respectively, higher and lower values after MCMC sampling. The ROS levels reached in Zhang's model are probably lower than in our cell system, so their formation was increased and their elimination decreased. We centered our GPx parameters' priors on the values used by Zhang *et al.* for GST. Since the observed GST and GPx omic data profiles were very different, it is not surprising that the posterior distributions of GPx parameters are clearly different from their prior. Through the Nrf2 pathway, CsA seems to have an important influence on GCLC and GCLM synthesis. While the basal transcription rates of GCLC and GCLM stabilize at values close to those of Zhang's *et al.*, parameters of GCLC and GCLM genes regulation by NMAs, linked to ROS and CsA levels, stabilize at values about four times higher.

The model gives access to unmeasured effects of CsA to cells, closer to a toxicity endpoint. The generation of ROS by CsA is an important toxicity mechanism for that molecule. The retro-control of ROS scavenging by ROS themselves, through the Nrf2 signaling pathway, induces a highly nonlinear behavior illustrated on Figure 6. ROS generation runs out of control at CsA exposure levels close to the high dose assayed *in vitro* (15 µM for extra-cellular concentration). We have an external validation of this finding. The 15 µM concentration was experimentally chosen to be the highest not affecting cell survival. We know that above that level, toxicity would start to have an impact on survival, so our model predictions seem reasonable. However, as in many systems biology models, only one signaling pathway has been taken into account in our model. Other ROS scavenging mechanisms are present in RPTECs and could be involved. On the other hand, CsA nephrotoxicity involves several mechanism [26 ,27 ,28 ,29 ,30] and it is possible that ROS generation is not enough to cause critical damages.

## Conclusion

Integrating omics approaches with mathematical systems biology models is still rarely done [31,32], even though that seems the best way to both understand the data and improve the predictive ability of the models [33 ,34]. Our modeling and simulations of the CsA mediated ROS production gives biologists insight into mechanisms of toxicity and provide quantitative estimates of toxicity beyond the time and dose range used in experiments. To go further, it would be interesting to have a more precise model description of GSH synthesis in the model, since cellular ROS concentrations are clearly correlated to GSH. It would also be interesting to couple this model with a physiological based pharmacokinetic (PBPK) model for CsA to be able to better predict human response.

# Tables

Table 1: *In vitro* CsA kinetic parameters description and their statistical distributions.

| Parameter | Description | Units | Prior | Posterior mode, mean ± SD |
|---|---|---|---|---|
| $CL_{in_1}$ | Diffusion rate constant for cellular uptake | $\mu m^3 \cdot sec^{-1}$ | $LU^*(10^{-1}, 10^4)$ | 99.6, 99.8 ± 21 |
| $Km_{out_1}$ | Michaelis constant for diffusion for cellular efflux | $zmol \cdot \mu m^{-3}$ | $LU(100, 50000)$ | 2965, 3160 ± 620 |
| $\dfrac{CL_{out_1}}{Km_{out_1}}$ | Diffusion rate constant over Michaelis constant for cellular efflux | $\mu m^3 \cdot sec^{-1}$ | $LU(10^{-2}, 20)$ | 0.581, 0.568 ± 0.16 |
| $k_1$ | Plastic binding rate constant | $sec^{-1}$ | $LU(10^{-6}, 5\times10^{-4})$ | $3.55\times10^{-5}$, $3.54\times10^{-5} \pm 1.0\times10^{-5}$ |
| $k_3$ | Power law coefficient for unbinding | dimensionless | Uniform(0, 0.95) | 0.921, 0.802 ± 0.074 |
| $k_2$ | Plastic unbinding rate constant | $zmol^{(1-k3)**} \cdot sec^{-1}$ | $LU(10^{-4}, 0.5)$ | $6.01\times10^{-4}$, $6.09\times10^{-3} \pm 8.7\times10^{-3}$ |
| $v_{max}$ | Maximum rate of metabolism | $zmol \cdot sec^{-1}$ | $LU(0.1, 5000)$ | 40.0, 47.2 ± 14 |
| $K_{m_2}$ | Michaelis constant for intra-cellular metabolism | zmol | $LU(5\cdot10^5, 5\cdot10^7)$ | $2.18\times10^6$, $3.43\times10^6 \pm 2.2\times10^6$ |

\**: *LU*: Log-uniform distribution (lower bound, upper bound).
\*: 1 zmol = 1 zeptomole = $10^{-21}$ mol

Table 2: Systems biology model parameters description and their statistical distributions.

| Parameter | Description | Units | Prior distribution | Posterior mode, mean ± SD |
|---|---|---|---|---|
| $v_{max\,7}$ | Maximum rate of CsA metabolism | $sec^{-1}$ | $LN(0.2, 3)$ | 0.187<br>0.274 ± 0.233 |
| $k_{f\,75}$ | Basal rate of ROS formation | $zmol.sec^{-1}$ | $LN(12, 3)$ | 79.1<br>135 ± 80.8 |
| $v_{max\,8b}$ | Maximum rate of ROS metabolism | $sec^{-1}$ | $LN(8, 3)$ | 2.67<br>3.88 ± 2.54 |
| $k_{ox\,10}$ | Keap1 oxidation rate constant | $zmol^{-1}.sec^{-1}$ | $Uniform(10^{-8}, 10^{-2})$ | $3.02 \times 10^{-6}$<br>$3.86 \times 10^{-6} \pm 2.72 \times 10^{-6}$ |
| $k_{ROS}$ | ROS formation rate constant | $sec^{-1}$ | $Uniform(10^{-8}, 10^{-2})$ | $6.55 \times 10^{-5}$<br>$8.86 \times 10^{-6} \pm 3.86 \times 10^{-5}$ |
| $k_{b\,18}$ | Nrf2 and Maf binding rate constant | $sec^{-1}$ | $LN(0.003, 3)$ | 0.0124<br>0.0193 ± 0.0167 |
| $k_{TSP\,21}$ | $mRNA_{CYP}$ transcription rate constant | $sec^{-1}$ | $LN(1.07, 3)$ | 1.29<br>1.65 ± 1.85 |
| $k_{TSP\,28}$ | $mRNA_{Nrf2}$ transcription rate constant | $sec^{-1}$ | $LN(0.00611, 3)$ | 0.087<br>0.062 ± 0.0603 |
| $k_{TSP\,34}$ | $mRNA_{GS}$ transcription rate constant | $sec^{-1}$ | $LN(1.15, 3)$ | 1.07<br>1.34 ± 0.53 |
| $k_{TSP\,42}$ | $mRNA_{GCLC}$ transcription rate constant | $sec^{-1}$ | $LN(1.98, 3)$ | 1.28<br>2.27 ± 1.91 |
| $k_{TSP\,48}$ | $mRNA_{GCLM}$ transcription rateconstant | $sec^{-1}$ | $LN(3.22, 3)$ | 3.95<br>4.84 ± 3.79 |
| $k_{TSP\,57}$ | $mRNA_{GST}$ transcription rate constant | $sec^{-1}$ | $LN(0.242, 3)$ | 0.021<br>0.553 ± 0.949 |
| $k_{TSP\,57b}$ | $mRNA_{GPx}$ transcription rate constant | $sec^{-1}$ | $LN(0.242, 3)$ | 0.098<br>0.123 ± 0.0779 |
| $k_{TSP\,66}$ | $mRNA_{MRP}$ transcription rate constant | $sec^{-1}$ | $LN(0.9, 3)$ | 1.22<br>2.23 ± 3.55 |
| $k_{b\,52}$ | GCLC and GCLM binding rate constant | $sec^{-1}$ | $LN(2 \times 10^{-5}, 3)$ | $4.33 \times 10^{-6}$<br>$1.09 \times 10^{-5} \pm 9.19 \times 10^{-6}$ |

[*]: 1 zmol = 1 zeptomole = $10^{-21}$ mol
[**]: *LN*: Log-normal distribution (mean, SD).

Table 2 (followed): Systems biology model parameters description and their statistical distributions.

| Parameter | Description | Posterior mode, mean ± SD | Prior distribution | Posterior mode, mean ± SD |
|---|---|---|---|---|
| $k_{ind(NMA)_{27}}$ | Induction coefficient for Nrf2 gene | zmol$^{-1}$.sec$^{-1}$ | $LN(100, 3)$ | 150<br>236 ± 433 |
| $k_{ind(NMA)_{33}}$ | Induction coefficient for GS gene | zmol$^{-1}$.sec$^{-1}$ | $LN(5.95, 3)$ | 2.17<br>3.85 ± 2.33 |
| $k_{ind(NMA)_{41}}$ | Induction coefficient for GCLC gene | zmol$^{-1}$.sec$^{-1}$ | $LN(8.7, 3)$ | 22.1<br>43.2 ± 25 |
| $k_{ind(NMA)_{47}}$ | Induction coefficient for GCLM gene | zmol$^{-1}$.sec$^{-1}$ | $LN(1.6, 3)$ | 3.28<br>5.75 ± 3.15 |
| $k_{ind(NMA)_{56}}$ | Induction coefficient for GST gene | zmol$^{-1}$.sec$^{-1}$ | $LN(11.9, 3)$ | 8.46<br>10.4 ± 8.61 |
| $k_{ind(NMA)_{56b}}$ | Induction coefficient for GPx gene | zmol$^{-1}$.sec$^{-1}$ | $LN(11.9, 3)$ | 1.37<br>6.75 ± 6.51 |
| $k_{ind(NMA)_{65}}$ | Induction coefficient for MRP gene | zmol$^{-1}$.sec$^{-1}$ | $LN(16, 3)$ | 6.43<br>9.62 ± 7.85 |
| $v_{max(GCL)_{72}}$ | Maximum rate of γ-GC synthesis | sec$^{-1}$ | $LN(8.2, 3)$ | 83.4<br>80.3 ± 67.5 |
| $v_{max(GCLC)_{72}}$ | Maximum rate of γ-GC synthesis | sec$^{-1}$ | $LN(1.9, 3)$ | 1.64<br>2.16 ± 3.11 |
| $v_{max_{73}}$ | Maximum rate of GSH synthesis | sec$^{-1}$ | $LN(6.5, 3)$ | 8.57<br>10.3 ± 4.43 |
| $v_{max_{74}}$ | Maximum rate of GSH degradation | zmol.sec$^{-1}$ | $LN(1845, 3)$ | 283<br>374 ± 353 |
| $Km_{74}$ | Michaelis-Menten constant of GSH degradation | zmol | $LN(2\times10^7, 3)$ | $1.62\times10^8$<br>$2.22\times10^8 \pm 2.36\times10^8$ |

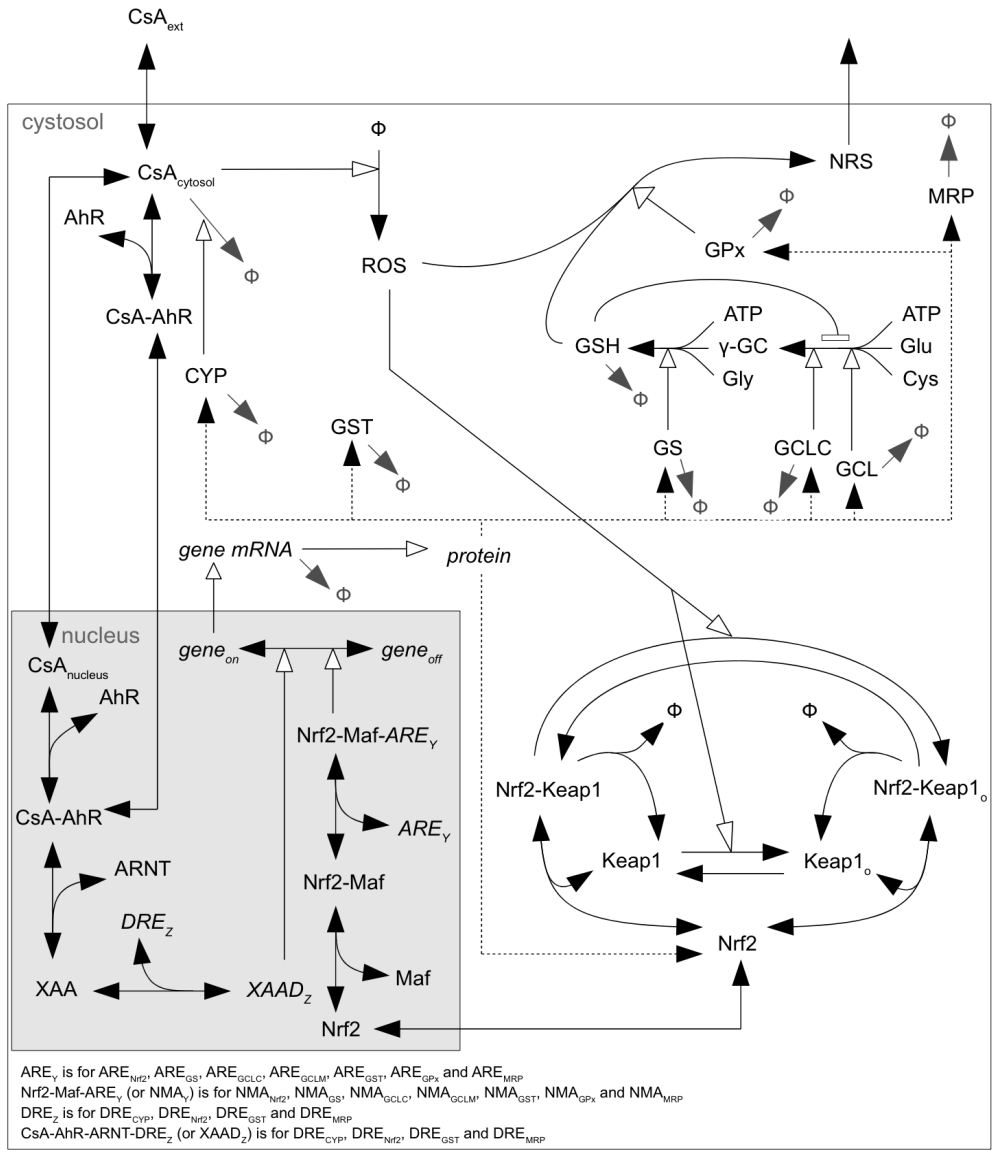

Figure 1: Schematic representation of the coupled PK biology systems model of the Nrf2 pathways.

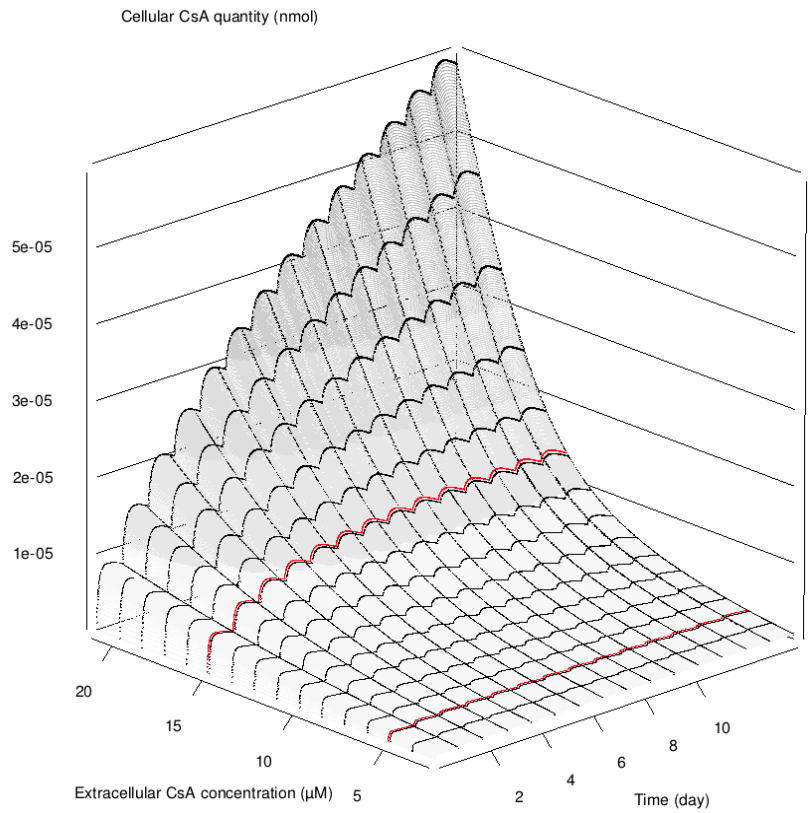

Figure 2: Predictions of intracellular CsA quantity (versus time and CsA administered). Thick red lines are predictions for 5 μM and 15 μM.

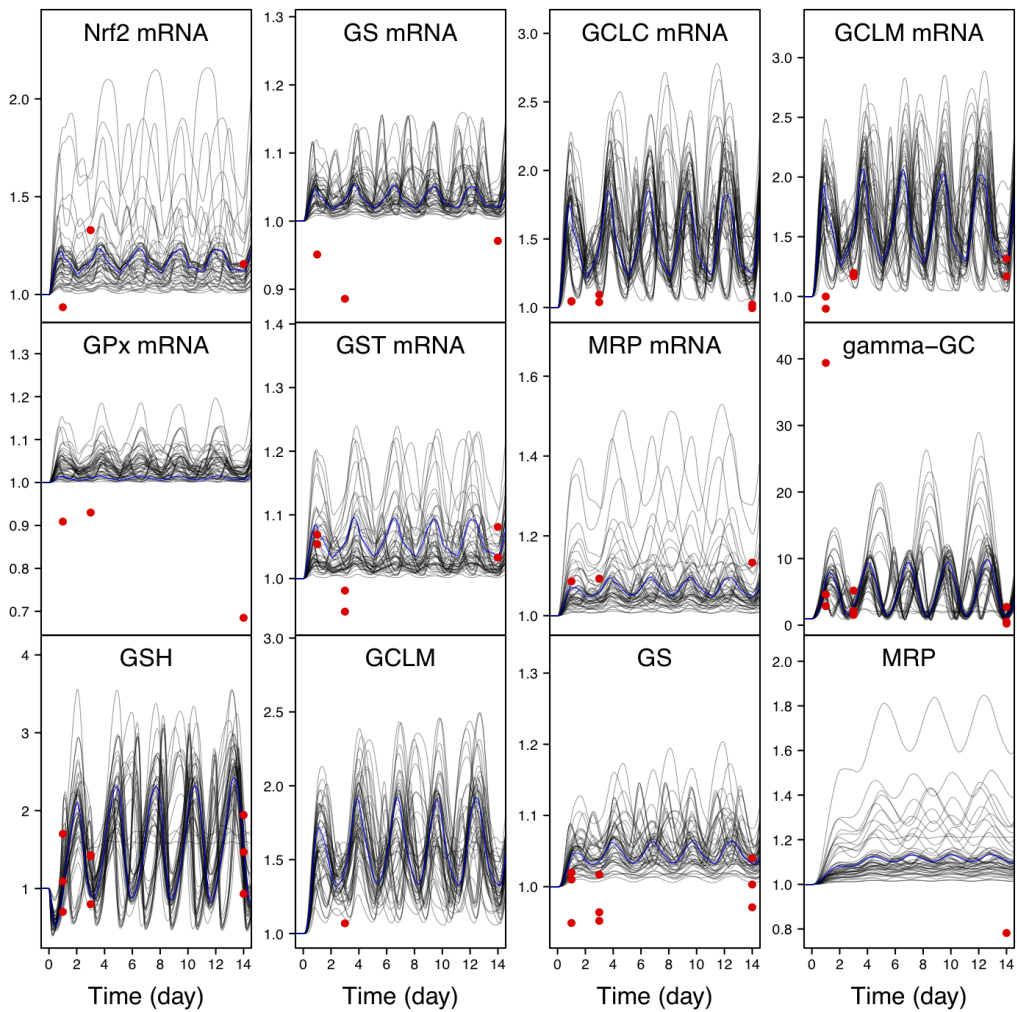

Figure 3: Omics fold-changes time-course in RPTEC cell with repeated low dose (5 μM) of CsA. Transcriptomics (Nrf2 mRNA, GS mRNA, GCLC mRNA, GCLM mRNA, GST mRNA, GPx mRNA and MRP mRNA) proteomics (GCLM, GS, and MRP), and metabolomics (γ-GC, and GSH) fold-changes time-course in RPTEC cells during 14 days with repeated low dose (5 μM) of CsA. The blue line indicates the best fitting (maximum posterior probability) model prediction. The black lines are predictions made with 49 random parameter sets. Red circles indicate data.

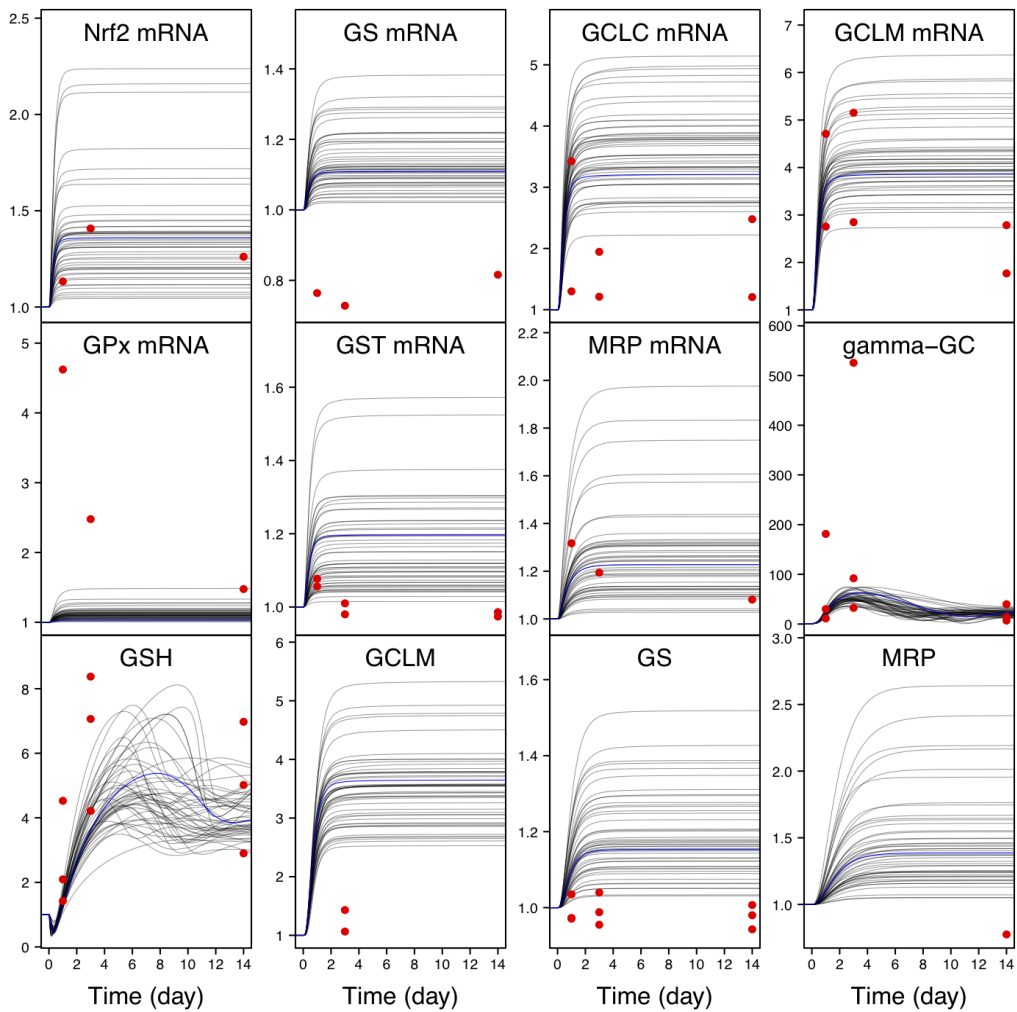

Figure 4: Omics fold-changes time-course in RPTEC cell with repeated high dose (15 µM) of CsA. Transcriptomics (Nrf2 mRNA, GS mRNA, GCLC mRNA, GCLM mRNA, GST mRNA, GPx mRNA and MRP mRNA) proteomics (GCLM, GS, and MRP), and metabolomics (γ-GC, and GSH) fold-changes time-course in RPTEC cells during 14 days with repeated high dose (15 µM) of CsA. The blue line indicates the best fitting (maximum posterior probability) model prediction. The black lines are predictions made with 49 random parameter sets. Red circles indicate data.

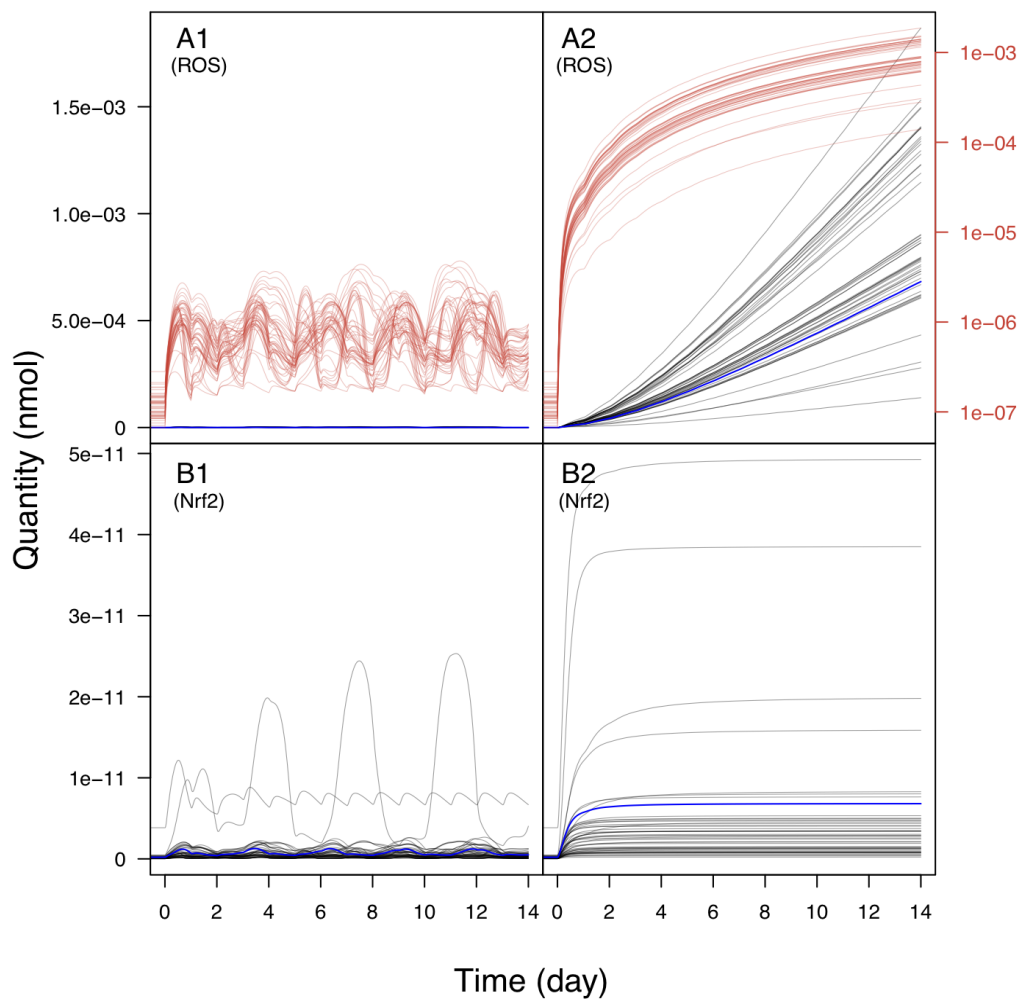

Figure 5: Pharmacokinetic modeling of cellular ROS and nuclear Nrf2 quantities. Pharmacokinetic modeling of cellular ROS quantity in cytosol for 5 μM (A1) and 15 μM (A2) and nuclear Nrf2 quantity for 5 μM (B1) and 15 μM (B2), during 14 days. The blue line indicates the best fitting (maximum posterior probability) model prediction. The black lines (normal scales) and red lines (semi-logarithmic scales) are predictions made with 49 random parameter sets.

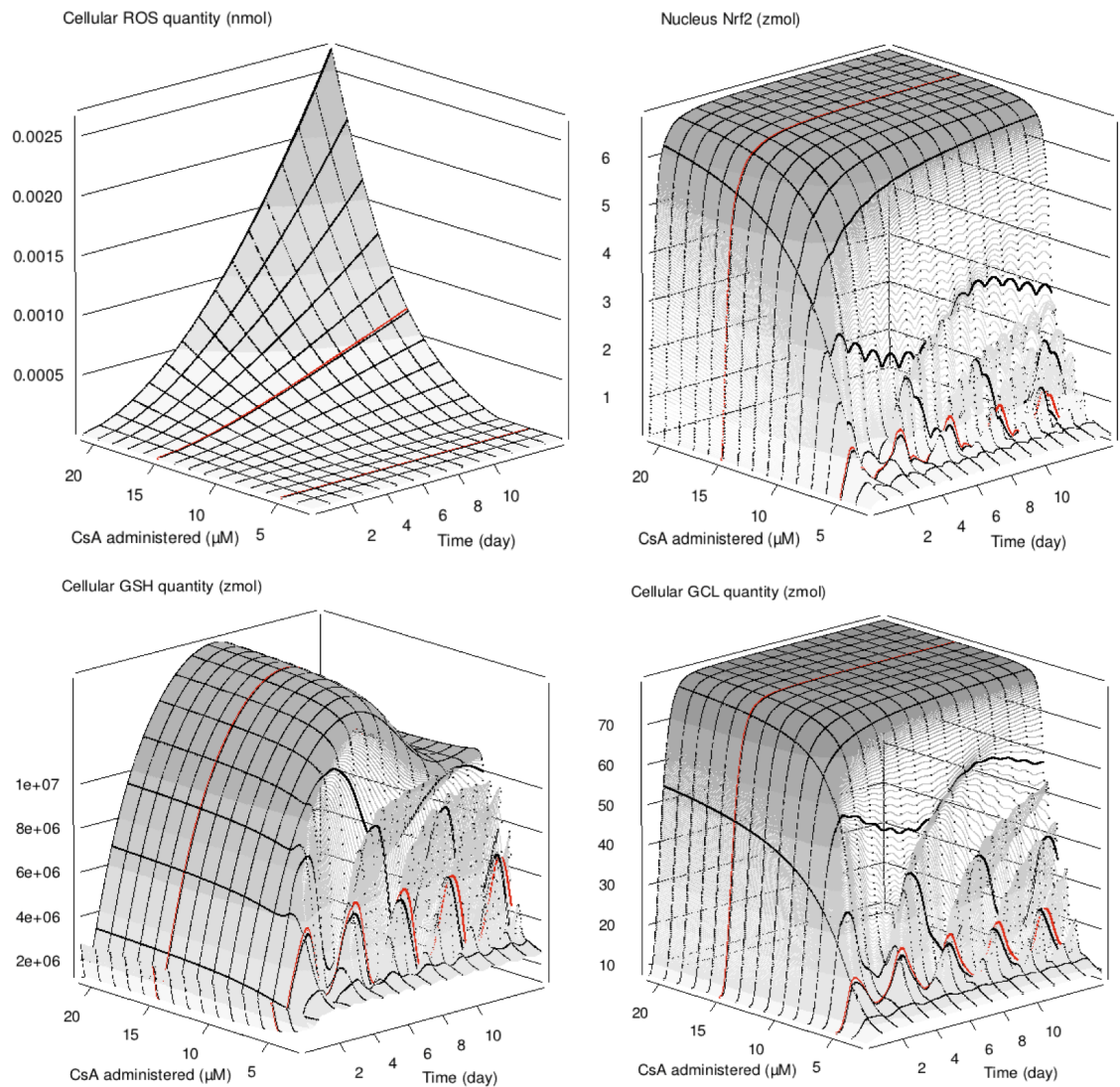

Figure 6: Predictions of cellular ROS, nuclear Nrf2, cellular GSH and cellular GCL quantities. Predictions of cellular ROS (top left), nuclear Nrf2 (top right), cellular GSH (down left) and cellular GCL (down right) quantities (versus time and dose). Thick red lines are predictions for 5 µM and 15 µM.

# Supplementary Material

**Differential equations of the Nrf2 model.**

**Table S1:** Model parameters values and initial state variables values.

**Table S2:** CsA quantities measured in the extracellular medium (3mL) at low CsA concentration exposure (5 μM).

**Table S3:** Cellular CsA quantities measured at low CsA concentration exposure (5 μM).

**Table S4:** CsA quantities measured on plastic at low CsA concentration exposure (5 μM).

**Table S5:** CsA quantities measured in the extracellular medium (3mL) at high CsA concentration exposure (15 μM).

**Table S6:** Cellular CsA quantities measured at high CsA concentration exposure (15 μM).

**Table S7:** CsA quantities measured on plastic at high CsA concentration exposure (15 μM).

**Table S8:** Fold changes measured at low CsA concentration (5 μM).

**Table S9:** Fold changes measured at high CsA concentration (15 μM).

**Figure S1**: Transcriptomics (Nrf2 mRNA, CYP mRNA, GS mRNA, GCLC mRNA, GCLM mRNA, GPx mRNA, and MRP mRNA) proteomics (GCLM, GS, and MRP), and metabolomics (γ-GC, and GSH) fold-changes time-course in RPTEC cells during 60 days with repeated low dose (5 μM) CsA dosing. The blue line indicates the best fitting (maximum posterior probability) model prediction. The black lines are predictions made with 49 random parameter sets.

**Figure S2:** Transcriptomics (Nrf2 mRNA, CYP mRNA, GS mRNA, GCLC mRNA, GCLM mRNA, GPx mRNA, and MRP mRNA) proteomics (GCLM, GS, and MRP), and metabolomics (γ-GC, and GSH) fold-changes time-course in RPTEC cells during 60 days with repeated high dose (15 μM) CsA dosing. The blue line indicates the best fitting (maximum posterior probability) model prediction. The black lines are predictions made with 49 random parameter sets.

**SUPPLEMENTARY MATERIAL**

**Differential equations of the Nrf2 model:**

$$\frac{\partial AhR_{cytosol}}{\partial t} = -k_{b_2} \cdot CsA_{cytosol} \cdot AhR_{cytosol} + k_{u_2} \cdot CsA\_AhR_{cytosol} \tag{A1}$$

$$\frac{\partial AhR_{nucleus}}{\partial t} = -k_{b_5} \cdot CsA_{nucleus} \cdot AhR_{nucleus} + k_{u_5} \cdot CsA\_AhR_{nucleus} \tag{A2}$$

$$\frac{\partial ARE_{GCLC}}{\partial t} = -k_{b_{40}} \cdot ARE_{GCLC} \left(Nrf2\_MAF_{nucleus}\right)^{n_{40}} + k_{u_{40}} \cdot NMA_{GCLC} \tag{A3}$$

$$\frac{\partial ARE_{GCLM}}{\partial t} = -k_{b_{46}} \cdot ARE_{GCLM} \left(Nrf2\_MAF_{nucleus}\right)^{n_{46}} + k_{u_{46}} \cdot NMA_{GCLM} \tag{A4}$$

$$\frac{\partial ARE_{GS}}{\partial t} = -k_{b_{32}} \cdot ARE_{GS} \left(Nrf2\_MAF_{nucleus}\right)^{n_{32}} + k_{u_{32}} \cdot NMA_{GS} \tag{A5}$$

$$\frac{\partial ARE_{GST}}{\partial t} \quad -k_{b_{54}} \cdot ARE_{GST} \left(Nrf2\_MAF_{nucleus}\right)^{n_{54}} + k_{u_{54}} \cdot NMA_{GST} \tag{A6}$$

$$\frac{\partial ARE_{GPx}}{\partial t} \quad -k_{b_{54b}} \cdot ARE_{GPx} \left(Nrf2\_MAF_{nucleus}\right)^{n_{54b}} + k_{u_{54b}} \cdot NMA_{GPx} \tag{A7}$$

$$\frac{\partial ARE_{MRP}}{\partial t} = -k_{b_{63}} \cdot ARE_{MRP} \left(Nrf2\_MAF_{nucleus}\right)^{n_{63}} + k_{u_{63}} \cdot NMA_{MRP} \tag{A8}$$

$$\frac{\partial ARE_{Nrf2}}{\partial t} = -k_{b_{25}} \cdot ARE_{Nrf2} \left(Nrf2\_MAF_{nucleus}\right)^{n_{25}} + k_{u_{25}} \cdot NMA_{Nrf2} \tag{A9}$$

$$\frac{\partial ARNT_{nucleus}}{\partial t} = -k_{b_6} \cdot CsA\_AhR_{nucleus} \cdot ARNT_{nucleus} + k_{u_6} \cdot XAA_{nucleus} \tag{A10}$$

$$\frac{\partial CsA_{cytosol}}{\partial t} = CL_{in_1} \cdot \frac{CsA_{extracellular}}{V_{extracellular}} - \frac{\frac{CL_{out_1}}{Km_{out_1}} \cdot \frac{CsA_{cytosol}}{V_{cytosol}}}{1 + \frac{CsA_{cytosol}}{V_{cytosol} \cdot Km_{out_1}}}$$

$$- \frac{v_{max_7} \cdot CYP_{cytosol} \cdot CsA_{cytosol}}{Km_2 + CsA_{cytosol}}$$

$$- k_{b_2} \cdot CsA_{cytosol} \cdot AhR_{cytosol} + k_{u_2} \cdot CsA\_AhR_{cytosol}$$

$$- CL_{in_4} \cdot \frac{CsA_{cytosol}}{V_{cytosol}} + CL_{out_4} \cdot \frac{CsA_{nucleus}}{V_{nucleus}}$$

(A11)

$$\frac{\partial CsA_{extracellular}}{\partial t} = -CL_{in_1} \cdot \frac{CsA_{extracellular}}{V_{extracellular}} + \frac{\frac{CL_{out_1}}{Km_{out_1}} \cdot \frac{CsA_{cytosol}}{V_{cytosol}}}{1 + \frac{CsA_{cytosol}}{V_{cytosol} \cdot Km_{out_1}}}$$

$$- k_1 \cdot CsA_{extracellular} + k_2 (CsA_{wall})^{k_3}$$

(A12)

$$\frac{\partial CsA_{nucleus}}{\partial t} = -k_{b_5} \cdot CsA_{nucleus} \cdot AhR_{nucleus} + k_{u_5} \cdot CsA\_AhR_{nucleus}$$

$$+ CL_{in_4} \cdot \frac{CsA_{cytosol}}{V_{cytosol}} - CL_{out_4} \cdot \frac{CsA_{nucleus}}{V_{nucleus}}$$

(A13)

$$\frac{\partial CsA_{wall}}{\partial t} = k_1 \cdot CsA_{extracellular} - k_2 (CsA_{wall})^{k_3}$$

(A14)

$$\frac{\partial CsA\_AhR_{cytosol}}{\partial t} = k_{b_2} \cdot CsA_{cytosol} \cdot AhR_{cytosol} - k_{u_2} \cdot CsA\_AhR_{cytosol}$$

$$- CL_{in_3} \cdot \frac{CsA\_AhR_{cytosol}}{V_{cytosol}} + CL_{out_3} \cdot \frac{CsA\_AhR_{nucleus}}{V_{nucleus}}$$

(A15)

$$\frac{\partial CsA\_AhR_{nucleus}}{\partial t} = CL_{in_3} \cdot \frac{CsA\_AhR_{cytosol}}{V_{cytosol}} - CL_{out_3} \cdot \frac{CsA\_AhR_{nucleus}}{V_{nucleus}}$$

$$+ k_{b_5} \cdot CsA_{nucleus} \cdot AhR_{nucleus} - k_{u_5} \cdot CsA\_AhR_{nucleus}$$

$$- k_{b_6} \cdot CsA\_AhR_{nucleus} \cdot ARNT_{nucleus} + k_{u_6} \cdot XAA_{nucleus}$$

(A16)

$$\frac{\partial CYP_{cytosol}}{\partial t} = -k_{deg_{24}} \cdot CYP_{cytosol} + k_{TSL_{23}} \cdot mRNA_{CYP}$$

(A17)

$$\frac{\partial DRE_{CYP}}{\partial t} = -k_{b_{19}} \cdot DRE_{CYP}(XAA_{nucleus})^{n_{19}} + k_{u_{19}} \cdot XAAD_{CYP} \quad (A18)$$

$$\frac{\partial DRE_{GST}}{\partial t} = -k_{b_{55}} \cdot DRE_{GST}(XAA_{nucleus})^{n_{55}} + k_{u_{55}} \cdot XAAD_{GST} \quad (A19)$$

$$\frac{\partial DRE_{MRP}}{\partial t} = -k_{b_{64}} \cdot DRE_{MRP}(XAA_{nucleus})^{n_{64}} + k_{u_{64}} \cdot XAAD_{MRP} \quad (A20)$$

$$\frac{\partial DRE_{Nrf2}}{\partial t} = -k_{b_{26}} \cdot DRE_{Nrf2}(XAA_{nucleus})^{n_{26}} + k_{u_{26}} \cdot XAAD_{Nrf2} \quad (A21)$$

$$\frac{\partial GCL_{cytosol}}{\partial t} = k_{b_{52}} \cdot GCLC_{cytosol} \cdot GCLM_{cytosol} - k_{u_{52}} \cdot GCL_{cytosol} \quad (A22)$$
$$- k_{deg_{53}} \cdot GCL_{cytosol}$$

$$\frac{\partial GCLC_{cytosol}}{\partial t} = -k_{b_{52}} \cdot GCLC_{cytosol} \cdot GCLM_{cytosol} + k_{u_{52}} \cdot GCL_{cytosol} \quad (A23)$$
$$- k_{deg_{45}} \cdot GCLC_{cytosol} + k_{TSL_{44}} \cdot mRNA_{GCLC}$$

$$\frac{\partial GCLM_{cytosol}}{\partial t} = -k_{b_{52}} \cdot GCLC_{cytosol} \cdot GCLM_{cytosol} + k_{u_{52}} \cdot GCL_{cytosol} \quad (A24)$$
$$- k_{deg_{51}} \cdot GCLM_{cytosol} + k_{TSL_{50}} \cdot mRNA_{GCLM}$$

$$\frac{\partial Gene_{CYP_{OFF}}}{\partial t} = -k_{act_{20}} \cdot Gene_{CYP_{OFF}} - k_{ind_{20}} \cdot Gene_{CYP_{OFF}} \cdot XAAD_{CYP} \quad (A25)$$
$$+ k_{desact_{20}} \cdot Gene_{CYP_{ON}}$$

$$\frac{\partial Gene_{CYP_{ON}}}{\partial t} = k_{act_{20}} \cdot Gene_{CYP_{OFF}} + k_{ind_{20}} \cdot Gene_{CYP_{OFF}} \cdot XAAD_{CYP} \quad (A25)$$
$$- k_{desact_{20}} \cdot Gene_{CYP_{ON}}$$

$$\frac{\partial Gene_{GCLC_{OFF}}}{\partial t} = -k_{act_{41}} \cdot Gene_{GCLC_{OFF}} - k_{ind_{41}} \cdot Gene_{GCLC_{OFF}} \cdot NMA_{GCLC} \quad (A27)$$
$$+ k_{desact_{41}} \cdot Gene_{GCCL_{ON}}$$

$$\frac{Gene_{GCLC_{ON}}}{\partial t} = k_{act_{41}} \cdot Gene_{GCLC_{OFF}} + k_{ind_{41}} \cdot Gene_{GCLC_{OFF}} \cdot NMA_{GCLC} \quad (A28)$$
$$- k_{desact_{41}} \cdot Gene_{GCLC_{ON}}$$

$$\frac{\partial Gene_{GCLM_{OFF}}}{\partial t} = -k_{act_{47}} \cdot Gene_{GCLM_{OFF}} - k_{ind_{47}} \cdot Gene_{GCLM_{OFF}} \cdot NMA_{GCLM} \quad (A29)$$
$$+ k_{desact_{47}} \cdot Gene_{GCLC_{ON}}$$

$$\frac{\partial Gene_{GCLM_{ON}}}{\partial t} = k_{act_{47}} \cdot Gene_{GCLM_{OFF}} + k_{ind_{47}} \cdot Gene_{GCLM_{OFF}} \cdot NMA_{GCLM} \quad (A30)$$
$$- k_{desact_{47}} \cdot Gene_{GCLM_{ON}}$$

$$\frac{\partial Gene_{GS_{OFF}}}{\partial t} = -k_{act_{33}} \cdot Gene_{GS_{OFF}} - k_{ind_{33}} \cdot Gene_{GS_{OFF}} \cdot NMA_{GS} \quad (A31)$$
$$+ k_{desact_{33}} \cdot Gene_{GS_{ON}}$$

$$\frac{\partial Gene_{GS_{ON}}}{\partial t} = k_{act_{33}} \cdot Gene_{GS_{OFF}} + k_{ind_{33}} \cdot Gene_{GS_{OFF}} \cdot NMA_{GS} \quad (A32)$$
$$- k_{desact_{33}} \cdot Gene_{GS_{ON}}$$

$$\frac{Gene_{GST_{OFF}}}{\partial t} = -k_{act_{56}} \cdot Gene_{GST_{OFF}} - k_{ind(NMA)_{56}} \cdot Gene_{GST_{OFF}} \cdot NMA_{GST} \quad (A33)$$
$$- k_{ind(XAAD)_{56}} \cdot Gene_{GST_{OFF}} \cdot XAAD_{GST} + k_{desact_{56}} \cdot Gene_{GST_{ON}}$$

$$\frac{Gene_{GST_{ON}}}{\partial t} = k_{act_{56}} \cdot Gene_{GST_{OFF}} + k_{ind(NMA)_{56}} \cdot Gene_{GST_{OFF}} \cdot NMA_{GST} \quad (A34)$$
$$k_{ind(XAAD)_{56}} \cdot Gene_{GST_{OFF}} \cdot XAAD_{GST} - k_{desact_{56}} \cdot Gene_{GST_{ON}}$$

$$\frac{Gene_{GPx_{OFF}}}{\partial t} = -k_{act_{56b}} \cdot Gene_{GPx_{OFF}} - k_{ind(NMA)_{56b}} \cdot Gene_{GPx_{OFF}} \cdot NMA_{GPx} \quad (A35)$$
$$+ k_{desact_{56b}} \cdot Gene_{GPx_{ON}}$$

$$\frac{Gene_{GPx_{ON}}}{\partial t} = k_{act_{56b}} \cdot Gene_{GPx_{OFF}} + k_{ind(NMA)_{56b}} \cdot Gene_{GPx_{OFF}} \cdot NMA_{GPx} \quad (A36)$$
$$- k_{desact_{56b}} \cdot Gene_{GPx_{ON}}$$

$$\frac{\partial Gene_{MRP_{OFF}}}{\partial t} = -k_{act_{65}} \cdot Gene_{MRP_{OFF}} - k_{ind(NMA)_{65}} \cdot Gene_{MRP_{OFF}} \cdot NMA_{MRP} \quad (A37)$$
$$- k_{ind(XAAD)_{65}} \cdot Gene_{MRP_{OFF}} \cdot XAAD_{MRP} + k_{desact_{65}} \cdot Gene_{MRP_{ON}}$$

$$\frac{\partial Gene_{MRP_{ON}}}{\partial t} = k_{act_{65}} \cdot Gene_{MRP_{OFF}} + k_{ind(NMA)_{65}} \cdot Gene_{MRP_{OFF}} \cdot NMA_{MRP} \quad \text{(A38)}$$
$$+ k_{ind(XAAD)_{65}} \cdot Gene_{MRP_{OFF}} \cdot XAAD_{MRP} - k_{desact_{65}} \cdot Gene_{MRP_{ON}}$$

$$\frac{\partial Gene_{Nrf2_{OFF}}}{\partial t} = -k_{act_{27}} \cdot Gene_{Nrf2_{OFF}} - k_{ind(NMA)_{27}} \cdot Gene_{Nrf2_{OFF}} \cdot NMA_{Nrf2}$$
$$- k_{ind(XAAD)_{27}} \cdot Gene_{Nrf2_{OFF}} \cdot XAAD_{Nrf2} + k_{desact_{27}} \cdot Gene_{Nrf2_{ON}}$$
(A39)

$$\frac{\partial Gene_{Nrf2_{ON}}}{\partial t} = k_{act_{27}} \cdot Gene_{Nrf2_{OFF}} + k_{ind(NMA)_{27}} \cdot Gene_{Nrf2_{OFF}} \cdot NMA_{Nrf2} \quad \text{(A40)}$$
$$+ k_{ind(XAAD)_{27}} \cdot Gene_{Nrf2_{OFF}} \cdot XAAD_{Nrf2} - k_{desact_{27}} \cdot Gene_{Nrf2_{ON}}$$

$$\frac{\partial GS_{cytosol}}{\partial t} = k_{b_{38}} \cdot GS\,mono_{cytosol} \cdot GS\,mono_{cytosol} - k_{u_{38}} \cdot GS_{cytosol} \quad \text{(A41)}$$
$$- k_{deg_{39}} \cdot GS_{cytosol}$$

$$\frac{\partial GS\,mono_{cytosol}}{\partial t} = -2 \cdot k_{b_{38}} \cdot GS\,mono_{cytosol} \cdot GS\,mono_{cytosol} + 2 \cdot k_{u_{38}} \cdot GS_{cytosol} \quad \text{(A42)}$$
$$- k_{deg_{37}} \cdot GS\,mono_{cytosol} + k_{TSL_{36}} \cdot mRNA_{GS}$$

$$R11 = \frac{ATP_{cytosol}}{K_{m(ATP)(GCL)_{72}}\left(1+\frac{GSH_{cytosol}}{K_{is(ATP)(GCL)_{72}}}\right) + ATP_{cytosol}\left(1+\frac{GSH_{cytosol}}{K_{ii(ATP)(GCL)_{72}}}\right)} \quad \text{(A43)}$$

$$R12 = \frac{Glu_{cytosol}}{K_{m(Glu)(GCL)_{72}}\left(1+\frac{GSH_{cytosol}}{K_{is(Glu)(GCL)_{72}}}\right) + Glu_{cytosol}\left(1+\frac{GSH_{cytosol}}{K_{ii(Glu)(GCL)_{72}}}\right)} \quad \text{(A44)}$$

$$R13 = \frac{v_{max(GCL)_{72}} \cdot GCL_{cytosol} \cdot Cys_{cytosol}}{K_{m(Cys)(GCL)_{72}} + Cys_{cytosol}} \quad \text{(A45)}$$

$$R21 = \frac{ATP_{cytosol}}{K_{m(ATP)(GCLC)_{72}}\left(1+\frac{GSH_{cytosol}}{K_{is(ATP)(GCLC)_{72}}}\right) + ATP_{cytosol}\left(1+\frac{GSH_{cytosol}}{K_{ii(ATP)(GCLC)_{72}}}\right)}$$
(A46)

$$R22 = \frac{Glu_{cytosol}}{K_{m(Glu)(GCLC)_{72}}\left(1+\frac{GSH_{cytosol}}{K_{is(Glu)(GCLC)_{72}}}\right)+Glu_{cytosol}\left(1+\frac{GSH_{cytosol}}{K_{ii(Glu)(GCLC)_{72}}}\right)} \quad (A46)$$

$$R23 = \frac{v_{max(GCLC)_{72}} \cdot GCLC_{cytosol} \cdot Cys_{cytosol}}{K_{m(Cys)(GCLC)_{72}}+Cys_{cytosol}} \quad (A48)$$

$$R31 = \frac{\dfrac{\gamma GC_{cytosol}}{K_{m1(\gamma GC)_{73}}}+\dfrac{\gamma GC^2_{cytosol}}{K_{m1(\gamma GC)_{73}} \cdot K_{m2(\gamma GC)_{73}}}}{1+\dfrac{2 \cdot \gamma GC_{cytosol}}{K_{m1(\gamma GC)_{73}}}+\dfrac{\gamma GC^2_{cytosol}}{K_{m1(\gamma GC)_{73}} \cdot K_{m2(\gamma GC)_{73}}}} \quad (A49)$$

$$R32 = \frac{v_{max_{73}} \cdot GS_{cytosol} \cdot Gly_{cytosol} \cdot ATP_{cytosol}}{\left(K_{m(Gly)_{73}}+Gly_{cytosol}\right)\left(K_{m(ATP)_{73}}+ATP_{cytosol}\right)} \quad (A50)$$

$$\frac{\partial \gamma GC_{cytosol}}{\partial t} = R11 \times R12 \times R13 + R21 \times R22 \times R23 - R31 \times R32 \quad (A51)$$

$$\frac{\partial GSH_{cytosol}}{\partial t} = -\frac{v_{max_{74}} \cdot GSH_{cytosol}}{K_{m_{74}}+GSH_{cytosol}} + R31 \times R32 \\ -\frac{v_{max_{8b}} \cdot GPx_{cytosol} \cdot GSH_{cytosol}}{K_{m(GSH)_{8b}}+GSH_{cytosol}} \cdot \frac{ROS_{cytosol}}{K_{m(ROS)_{8b}}+ROS_{cytosol}} \quad (A52)$$

$$\frac{\partial GST_{cytosol}}{\partial t} = k_{b_{61}} \cdot GST\,mono_{cytosol} \cdot GST\,mono_{cytosol} - k_{u_{61}} \cdot GST_{cytosol} \\ - k_{deg_{62}} \cdot GST_{cytosol} \quad (A53)$$

$$\frac{\partial GST\,mono_{cytosol}}{\partial t} = -2.k_{b_{61}} \cdot GST\,mono_{cytosol} \cdot GST\,mono_{cytosol} + 2.k_{u_{61}} \cdot GST_{cytosol} \\ -k_{deg_{60}} \cdot GST\,mono_{cytosol} + k_{TSL_{59}} \cdot mRNA_{GST} \quad (A54)$$

$$\frac{\partial Keap1_{cytosol}}{\partial t} = k_{red_{10}} \cdot Keap1o_{cytosol} - k_{ox_{10}} \cdot Keap1_{cytosol} \cdot ROS_{cytosol} \\ +k_{u_{14}} \cdot Nrf2\_Keap1_{cytosol} - k_{b_{14}} \cdot Nrf2_{cytosol} \cdot Keap1_{cytosol} \\ +k_{u_{12}} \cdot Nrf2\_Keap1_{cytosol} \quad (A55)$$

$$\frac{\partial Keap1o_{cytosol}}{\partial t} = \begin{aligned} &-k_{red_{10}}.Keap1o_{cytosol}+k_{ox_{10}}.Keap1_{cytosol}.ROS_{cytosol}\\ &+k_{u_{15}}.Nrf2\_Keap1o_{cytosol}-k_{b_{15}}.Nrf2_{cytosol}.Keap1o_{cytosol}\\ &+k_{u_{13}}.Nrf2\_Keap1o_{cytosol} \end{aligned} \quad (A56)$$

$$\frac{\partial mRNA_{CYP}}{\partial t} = -k_{deg_{22}}.mRNA_{CYP}+k_{TSP_{21}}.Gene_{CYP_{ON}} \quad (A57)$$

$$\frac{\partial mRNA_{GCLC}}{\partial t} = -k_{deg_{43}}.mRNA_{GCLC}+k_{TSP_{42}}.Gene_{GCLC_{ON}} \quad (A58)$$

$$\frac{\partial mRNA_{GCLM}}{\partial t} = -k_{deg_{49}}.mRNA_{GCLM}+k_{TSP_{48}}.Gene_{GCLM_{ON}} \quad (A59)$$

$$\frac{\partial mRNA_{GS}}{\partial t} = -k_{deg_{35}}.mRNA_{GS}+k_{TSP_{34}}.Gene_{GS_{ON}} \quad (A60)$$

$$\frac{\partial mRNA_{GST}}{\partial t} = -k_{deg_{58}}.mRNA_{GST}+k_{TSP_{57}}.Gene_{GST_{ON}} \quad (A61)$$

$$\frac{\partial mRNA_{GPx}}{\partial t} = -k_{deg_{58b}}.mRNA_{GPx}+k_{TSP_{57b}}.Gene_{GPx_{ON}} \quad (A62)$$

$$\frac{\partial mRNA_{MRP}}{\partial t} = -k_{deg_{67}}.mRNA_{MRP}+k_{TSP_{66}}.Gene_{MRP_{ON}} \quad (A63)$$

$$\frac{\partial mRNA_{Nrf2}}{\partial t} = -k_{deg_{29}}.mRNA_{Nrf2}+k_{TSP_{28}}.Gene_{Nrf2_{ON}} \quad (A64)$$

$$\frac{\partial MRP_{cytosol}}{\partial t} = \begin{aligned} &k_{b_{70}}.MRP\,mono_{cytosol}.MRP\,mono_{cytosol}-k_{u_{70}}.MRP_{cytosol}\\ &-k_{deg_{71}}.MRP_{cytosol} \end{aligned} \quad (A65)$$

$$\frac{\partial MRP\,mono_{cytosol}}{\partial t} = \begin{aligned} &-2.k_{b_{70}}.MRP\,mono_{cytosol}.MRP\,mono_{cytosol}+2.k_{u_{70}}.MRP_{cytosol}\\ &-k_{deg_{69}}.MRP\,mono_{cytosol}+k_{TSL_{68}}.mRNA_{MRP} \end{aligned}$$
(A66)

$$\frac{\partial MAF_{nucleus}}{\partial t} = -k_{b_{18}}.MAF_{nucleus}.Nrf2_{nucleus}+k_{u_{18}}.Nrf2\_MAF_{nucleus} \quad (A67)$$

$$\frac{\partial NMA_{GCLC}}{\partial t} = k_{b_{40}}.ARE_{GCLC}\left(Nrf2\_MAF_{nucleus}\right)^{n_{40}}-k_{u_{40}}.NMA_{GCLC} \quad (A68)$$

$$\frac{\partial NMA_{GCLM}}{\partial t} = k_{b_{46}} \cdot ARE_{GCLM}(Nrf2\_MAF_{nucleus})^{n_{46}} - k_{u_{46}} \cdot NMA_{GCLM} \tag{A69}$$

$$\frac{\partial NMA_{GS}}{\partial t} = k_{b_{32}} \cdot ARE_{GS}(Nrf2\_MAF_{nucleus})^{n_{32}} - k_{u_{32}} \cdot NMA_{GS} \tag{A70}$$

$$\frac{\partial NMA_{GST}}{\partial t} = k_{b_{54}} \cdot ARE_{GST}(Nrf2\_MAF_{nucleus})^{n_{54}} - k_{u_{54}} \cdot NMA_{GST} \tag{A71}$$

$$\frac{\partial NMA_{GPx}}{\partial t} = k_{b_{54b}} \cdot ARE_{GPx}(Nrf2\_MAF_{nucleus})^{n_{54b}} - k_{u_{54b}} \cdot NMA_{GPx} \tag{A72}$$

$$\frac{\partial NMA_{MRP}}{\partial t} = k_{b_{63}} \cdot ARE_{MRP}(Nrf2\_MAF_{nucleus})^{n_{63}} - k_{u_{63}} \cdot NMA_{MRP} \tag{A73}$$

$$\frac{\partial NMA_{Nrf2}}{\partial t} = k_{b_{25}} \cdot ARE_{Nrf2}(Nrf2\_MAF_{nucleus})^{n_{25}} - k_{u_{25}} \cdot NMA_{Nrf2} \tag{A74}$$

$$\begin{aligned}\frac{\partial Nrf2_{cytosol}}{\partial t} = & -k_{deg_{31}} \cdot Nrf2_{cytosol} + k_{TSL_{30}} \cdot mRNA_{Nrf2} \\ & -k_{b_{15}} \cdot Nrf2_{cytosol} \cdot Keap1o_{cytosol} + k_{u_{15}} \cdot Nrf2\_Keap1o_{cytosol} \\ & -k_{b_{14}} \cdot Nrf2_{cytosol} \cdot Keap1_{cytosol} + k_{u_{14}} \cdot Nrf2\_Keap1_{cytosol} \\ & -CL_{in_{16}} \cdot \frac{Nrf2_{cytosol}}{V_{cytosol}} + CL_{out_{16}} \cdot \frac{Nrf2_{nucleus}}{V_{nucleus}} \end{aligned} \tag{A75}$$

$$\begin{aligned}\frac{\partial Nrf2_{nucleus}}{\partial t} = & -k_{b_{18}} \cdot Nrf2_{nucleus} \cdot Maf_{nucleus} + k_{u_{18}} \cdot Nrf2\_Maf_{nucleus} \\ & +CL_{in_{16}} \cdot \frac{Nrf2_{cytosol}}{V_{cytosol}} - CL_{out_{16}} \cdot \frac{Nrf2_{nucleus}}{V_{nucleus}} \\ & -k_{deg_{17}} \cdot Nrf2_{nucleus} \end{aligned} \tag{A76}$$

$$\begin{aligned}\frac{\partial Nrf2\_Keap1_{cytosol}}{\partial t} = & \ k_{red_{11}} \cdot Nrf2\_Keap1o_{cytosol} \\ & -k_{ox_{11}} \cdot Nrf2\_Keap1_{cytosol} \cdot ROS_{cytosol} \\ & -k_{u_{14}} \cdot Nrf2\_Keap1_{cytosol} \\ & +k_{b_{14}} \cdot Nrf2_{cytosol} \cdot Keap1_{cytosol} \\ & -k_{u_{12}} \cdot Nrf2\_Keap1_{cytosol} \end{aligned} \tag{A77}$$

$$\frac{\partial Nrf2\_Keap1o_{cytosol}}{\partial t} = -k_{red_{11}} \cdot Nrf2\_Keap1o_{cytosol}$$
$$+ k_{ox_{11}} \cdot Nrf2\_Keap1_{cytosol} \cdot ROS_{cytosol}$$
$$- k_{u_{15}} \cdot Nrf2\_Keap1o_{cytosol} \quad (A78)$$
$$+ k_{b_{15}} \cdot Nrf2_{cytosol} \cdot Keap1o_{cytosol}$$
$$- k_{u_{13}} \cdot Nrf2\_Keap1o_{cytosol}$$

$$\frac{\partial Nrf2\_Maf_{nucleus}}{\partial t} = -2 \cdot k_{b_{54}} \cdot ARE_{GST} (Nrf2\_MAF_{nucleus})^{n_{54}} + 2 \cdot k_{u_{54}} \cdot NMA_{GST}$$
$$- 2 \cdot k_{b_{54b}} \cdot ARE_{GPx} (Nrf2\_MAF_{nucleus})^{n_{54b}} + 2 \cdot k_{u_{54b}} \cdot NMA_{GPx}$$
$$- 3 \cdot k_{b_{46}} \cdot ARE_{GCLM} (Nrf2\_MAF_{nucleus})^{n_{46}} + 3 \cdot k_{u_{46}} \cdot NMA_{GCLM}$$
$$- k_{b_{25}} \cdot ARE_{Nrf2} (Nrf2\_MAF_{nucleus})^{n_{25}} + k_{u_{25}} \cdot NMA_{Nrf2} \quad (A79)$$
$$- 2 \cdot k_{b_{63}} \cdot ARE_{MRP} \cdot (Nrf2\_MAF_{nucleus})^{n_{63}} + 2 \cdot k_{u_{63}} \cdot NMA_{MRP}$$
$$- 3 \cdot k_{b_{40}} \cdot ARE_{GCLC} (Nrf2\_MAF_{nucleus})^{n_{40}} + 3 \cdot k_{u_{40}} \cdot NMA_{GCLC}$$
$$- 2 \cdot k_{b_{32}} \cdot ARE_{GS} (Nrf2\_MAF_{nucleus})^{n_{32}} + 2 \cdot k_{u_{32}} \cdot NMA_{GS}$$
$$+ k_{b_{18}} \cdot Nrf2_{nucleus} \cdot MAF_{nucleus} - k_{u_{18}} \cdot Nrf2\_MAF_{nuleus}$$

$$\frac{\partial ROS_{cytosol}}{\partial t} = k_{f_{75}} + k_{ROS} \cdot CsA_{cytosol}$$
$$- \frac{v_{max_{8b}} \cdot GPx_{cytosol} \cdot GSH_{cytosol}}{K_{m(GSH)_{8b}} + GSH_{cytosol}} \cdot \frac{ROS_{cytosol}}{K_{m(ROS)_{8b}} + ROS_{cytosol}} \quad (A80)$$

$$\frac{\partial XAA_{nucleus}}{\partial t} = -2 \cdot k_{b_{26}} \cdot DRE_{Nrf2} (XAA_{nucleus})^{n_{26}} + 2 \cdot k_{u_{26}} \cdot XAAD_{Nrf2}$$
$$- 2 \cdot k_{b_{19}} \cdot DRE_{CYP} (XAA_{nucleus})^{n_{19}} + 2 \cdot k_{u_{19}} \cdot XAAD_{CYP}$$
$$- 2 \cdot k_{b_{55}} \cdot DRE_{GST} (XAA_{nucleus})^{n_{55}} + 2 \cdot k_{u_{55}} \cdot XAAD_{GST} \quad (A81)$$
$$- k_{b_{64}} \cdot DRE_{MRP} (XAA_{nucleus})^{n_{64}} + k_{u_{64}} \cdot XAAD_{MRP}$$
$$+ k_{b_6} \cdot X\_AhR_{nucleus} \cdot ARNT_{nucleus} - k_{u_6} \cdot XAA_{nucleus}$$

$$\frac{\partial XAAD_{CYP}}{\partial t} = k_{b_{19}} \cdot DRE_{CYP} (XAA_{nucleus})^{n_{19}} - k_{u_{19}} \cdot XAAD_{CYP} \quad (A82)$$

$$\frac{\partial XAAD_{GST}}{\partial t} = k_{b_{55}} \cdot DRE_{GST} \left(XAA_{nucleus}\right)^{n_{55}} - k_{u_{55}} \cdot XAAD_{GST} \qquad (A83)$$

$$\frac{\partial XAAD_{MRP}}{\partial t} = k_{b_{64}} \cdot DRE_{MRP} \left(XAA_{nucleus}\right)^{n_{64}} - k_{u_{64}} \cdot XAAD_{MRP} \qquad (A84)$$

$$\frac{\partial XAAD_{Nrf2}}{\partial t} = k_{b_{26}} \cdot DRE_{Nrf2} \left(XAA_{nucleus}\right)^{n_{26}} - k_{u_{26}} \cdot XAAD_{Nrf2} \qquad (A85)$$

Table S1: Model parameters values and initial state variables values.

| Equation | Parameters and initial state variables | | |
|---|---|---|---|
| A1: $\frac{\partial AhR_{cytosol}}{\partial t}$ | $AhR_{cytosol_{(initial)}}$ = 34 zmol | $k_{b_2}$ = 0 zmol$^{-1}$.s$^{-1}$ | $k_{u_2}$ = 0.02 s$^{-1}$ |
| A2: $\frac{\partial AhR_{nucleus}}{\partial t}$ | $AhR_{nucleus_{(initial)}}$ = 0 zmol | $k_{b_5}$ = 0 zmol$^{-1}$.s$^{-1}$ | $k_{u_5}$ = 0.02 s$^{-1}$ |
| A3: $\frac{\partial ARE_{GCLC}}{\partial t}$ | $k_{b_{40}}$ = 14.6 zmol$^{-3}$.s$^{-1}$ $\quad$ $ARE_{GCLC_{(initial)}}$ = 6.35E$^{-4}$ zmol | $k_{u_{40}}$ = 0.02 s$^{-1}$ | $n_{40}$ = 3 |
| A4: $\frac{\partial ARE_{GCLM}}{\partial t}$ | $k_{b_{46}}$ = 14.6 zmol$^{-3}$.s$^{-1}$ $\quad$ $ARE_{GCLM_{(initial)}}$ = 6.35.E$^{-4}$ zmol | $k_{u_{46}}$ = 0.02 s$^{-1}$ | $n_{46}$ = 3 |
| A5: $\frac{\partial ARE_{GS}}{\partial t}$ | $k_{b_{32}}$ = 0.29 zmol$^{-2}$.s$^{-1}$ $\quad$ $ARE_{GS_{(initial)}}$ = 6.35E$^{-4}$ zmol | $k_{u_{32}}$ = 0.02 s$^{-1}$ | $n_{32}$ = 2 |
| A6: $\frac{\partial ARE_{GST}}{\partial t}$ | $k_{b_{54}}$ = 0.29 zmol$^{-2}$.s$^{-1}$ $\quad$ $ARE_{GST_{(initial)}}$ = 6.35E$^{-4}$ zmol | $k_{u_{54}}$ = 0.02 s$^{-1}$ | $n_{54}$ = 2 |
| A7: $\frac{\partial ARE_{GPx}}{\partial t}$ | $k_{b_{54b}}$ = 0.29 zmol$^{-2}$.s$^{-1}$ $\quad$ $ARE_{GPx_{(initial)}}$ = 6.35E$^{-4}$ zmol | $k_{u_{54b}}$ = 0.02 s$^{-1}$ | $n_{54b}$ = 2 |
| A8: $\frac{\partial ARE_{MRP}}{\partial t}$ | $k_{b_{63}}$ = 0.29 zmol$^{-2}$.s$^{-1}$ $\quad$ $ARE_{MRP_{(initial)}}$ = 6.35E$^{-4}$ zmol | $k_{u_{63}}$ = 0.02 s$^{-1}$ | $n_{63}$ = 2 |
| A9: $\frac{\partial ARE_{Nrf2}}{\partial t}$ | $k_{b_{25}}$ = 5.26E$^{-3}$ zmol$^{-1}$.s$^{-1}$ $\quad$ $ARE_{Nrf2_{(initial)}}$ = 6.35E$^{-3}$ zmol | $k_{u_{25}}$ = 0.02 s$^{-1}$ | $n_{25}$ = 1 |
| A10: $\frac{\partial ARNT_{Nucleus}}{\partial t}$ | $k_{b_6}$ = 0.014 zmol$^{-1}$.s$^{-1}$ | $k_{u_6}$ = 0.002 s$^{-1}$ | $ARNT_{(initial)}$ = 6.35E$^{-3}$ zmol |

| Equation | Parameters and initial state variables | | |
|---|---|---|---|
| A11: $\frac{\partial CsA_{cytosol}}{\partial t}$ | $V_{cytosol}$ = 1702 μm³ | $CL_{in_1}$ = 99.6 zmol.sec⁻¹ | $CL_{out_1}$ = 1483 zmol.sec⁻¹ |
| | $V_{nucleus}$ = 380 μm³ | $Km_{out_1}$ = 2965 μmol.L⁻¹ | $k_3$ = 0.921 |
| A12: $\frac{\partial CsA_{extracellar}}{\partial t}$ | $CsA_{cytosol_{(initial)}}$ = 0 zmol | $V_{max_2}$ = 0.2 sec⁻¹ | $Km_2$ = 2.18E⁶ zmol |
| | $CsA_{wall_{(initial)}}$ = 0 zmol | $CL_{in_4}$ = $CL_{in_1}$ | $CL_{out_4}$ = $CL_{out_1}$ |
| A13: $\frac{\partial CsA_{nuleus}}{\partial t}$ | $CsA_{nucleus_{(initial)}}$ = 0 zmol | $k_1$ = 3.55E⁻⁵ sec⁻¹ | |
| A14: $\frac{\partial CsA_{wall}}{\partial t}$ | $k_2$ = 6.01E⁻⁴ zmol$^{(1-k3)}$.sec⁻¹ | Refer to A1 and A2 for the other parameters. | |
| A15: $\frac{\partial CsA\_AhR_{cytosol}}{\partial t}$ | $CsA\_AhR_{cytosol_{(initial)}}$ = 0 zmol | $CL_{in_3}$ = 10 μm³.s⁻¹ | $CL_{out_3}$ = 1 μm³.sec⁻¹ |
| | Refer to A1 for the other parameters. | | |
| A16: $\frac{\partial CsA\_AhR_{nucleus}}{\partial t}$ | $CsA\_AhR_{nucleus_{(initial)}}$ = 0 zmol | Refer to A2, A11 and A15 for the parameters. | |
| A17: $\frac{\partial CYP_{cytosol}}{\partial t}$ | $CYP_{cytosol_{(initial)}}$ = 34 zmol | $k_{deg_{24}}$ = 1E-4 s⁻¹ | $k_{TSL_{23}}$ = 0041 s⁻¹ |
| A18: $\frac{\partial DRE_{CYP}}{\partial t}$ | $k_{b_{19}}$ = 1.39 zmol⁻².s⁻¹ | $k_{u_{19}}$ = 0.018 s⁻¹ | $n_{19}$ = 2 |
| | $DRE_{CYP_{(initial)}}$ = 6.35E⁻⁴ zmol | | |
| A19: $\frac{\partial DRE_{GST}}{\partial t}$ | $k_{b_{55}}$ = 13.9 zmol⁻².s⁻¹ | $k_{u_{55}}$ = 0.02 s⁻¹ | $n_{55}$ = 2 |
| | $DRE_{GST_{(initial)}}$ = 6.35E⁻⁴ zmol | | |
| A20: $\frac{\partial DRE_{MRP}}{\partial t}$ | $k_{b_{64}}$ = 0.26 zmol⁻².s⁻¹ | $k_{u_{64}}$ = 0.02 s⁻¹ | $n_{64}$ = 1 |
| | $DRE_{MRP_{(initial)}}$ = 6.35E⁻⁴ zmol | | |
| A21: $\frac{\partial DRE_{Nrf2}}{\partial t}$ | $k_{b_{26}}$ = 13.9 zmol⁻².s⁻¹ | $k_{u_{26}}$ = 0.02 s⁻¹ | $n_{26}$ = 2 |
| | $DRE_{Nrf2_{(initial)}}$ = 6.35E⁻⁴ zmol | | |

| Equation | Parameters and initial state variables | | |
|---|---|---|---|
| A22: $\frac{\partial GCL_{cytosol}}{\partial t}$ | $k_{b_{52}}$ = -[a] | $k_{u_{52}}$ = 0.02 s$^{-1}$ | $k_{deg_{53}}$ = 3.86E-5 s$^{-1}$ |
| | $GCL_{cytosol_{(initial)}}$ = 1020 zmol | | |
| A23: $\frac{\partial GCLC_{cytosol}}{\partial t}$ | $GCLC_{cytosol_{(initial)}}$ = 2890 zmol | $k_{deg_{45}}$ = 3.86E-5 s$^{-1}$ | $k_{TSL_{44}}$ = 0.0417 s$^{-1}$ |
| | Refer to A22 for the other parameters. | | |
| A24: $\frac{\partial GCLM_{cytosol}}{\partial t}$ | $GCLM_{cytosol_{(initial)}}$ = 612 zmol | $k_{deg_{51}}$ = 3.86E-5 s$^{-1}$ | $k_{TSL_{50}}$ = 0.0417 s$^{-1}$ |
| | Refer to A22 for the other parameters. | | |
| A25: $\frac{\partial Gene_{CYP_{OFF}}}{\partial t}$ | $Gene_{CYP_{OFF_{(initial)}}}$ = 6.35E-4 zmol | $k_{act_{20}}$ = 2.5E-5 s$^{-1}$ | $k_{desact_{20}}$ = 0.01 s$^{-1}$ |
| A26: $\frac{\partial Gene_{CYP_{ON}}}{\partial t}$ | $Gene_{CYP_{ON_{(initial)}}}$ = 0 zmol | $k_{ind(NMA)_{20}}$ = 0.079 zmol$^{-1}$.s$^{-1}$ | |
| A27: $\frac{\partial Gene_{GCLC_{OFF}}}{\partial t}$ | $Gene_{GCLC_{OFF_{(initial)}}}$ = 6.35E-4 zmol | $k_{act_{41}}$ = 4E-4 s$^{-1}$ | $k_{desact_{41}}$ = 0.01 s$^{-1}$ |
| 28: $\frac{\partial Gene_{GCLC_{ON}}}{\partial t}$ | $Gene_{GCLC_{ON_{(initial)}}}$ = 0 zmol | $k_{ind_{41}}$ = -[a] | |
| A29: $\frac{\partial Gene_{GCLM_{OFF}}}{\partial t}$ | $Gene_{GCLM_{OFF_{(initial)}}}$ = 6.35E-4 zmol | $k_{act_{47}}$ = 4E-4 s$^{-1}$ | $k_{desact_{47}}$ = 0.01 s$^{-1}$ |
| A30: $\frac{\partial Gene_{GCLM_{ON}}}{\partial t}$ | $Gene_{GCLM_{ON_{(initial)}}}$ = 0 zmol | $k_{ind(NMA)_{47}}$ = -[a] | |
| A31: $\frac{\partial Gene_{GS_{OFF}}}{\partial t}$ | $Gene_{GS_{OFF_{(initial)}}}$ = 6.35E-4 zmol | $k_{act_{33}}$ = 5E-4 s$^{-1}$ | $k_{desact_{33}}$ = 0.01 s$^{-1}$ |
| A32: $\frac{\partial Gene_{GS_{ON}}}{\partial t}$ | $Gene_{GS_{ON_{(initial)}}}$ = 0 zmol | $k_{ind(NMA)_{33}}$ = -[a] | |
| A33: $\frac{\partial Gene_{GST_{OFF}}}{\partial t}$ | $Gene_{GST_{OFF_{(initial)}}}$ = 6.35E-4 zmol | $k_{act_{56}}$ = 1E-3 s$^{-1}$ | $k_{desact_{56}}$ = 0.01 s$^{-1}$ |
| A34: $\frac{\partial Gene_{GST_{ON}}}{\partial t}$ | $Gene_{GST_{ON_{(initial)}}}$ = 0 zmol | $k_{ind(NMA)_{56}}$ = -[a] | $k_{ind(XAAD)_{56}}$ = 0.26 zmol$^{-1}$.s$^{-1}$ |

[a] : parameters estimated.

| Equation | Parameters and initial state variables | | |
|---|---|---|---|
| A35: $\frac{\partial Gene_{GPx_{OFF}}}{\partial t}$ | $Gene_{GPx_{OFF}(initial)}$ = 6.35E$^{-4}$ zmol | $k_{act_{56b}}$ = 1E$^{-3}$ s$^{-1}$ | $k_{desact_{56b}}$ = 0.01 s$^{-1}$ |
| A36: $\frac{\partial Gene_{GPx_{ON}}}{\partial t}$ | $Gene_{GPx_{ON}(initial)}$ = 0 zmol | $k_{ind(NMA)_{56b}}$ = -$^a$ | |
| A37: $\frac{\partial Gene_{MRP_{OFF}}}{\partial t}$ | $Gene_{MRP_{OFF}(initial)}$ = 6.35E$^{-4}$ zmol | $k_{act_{65}}$ = 6.8E$^{-4}$ s$^{-1}$ | $k_{desact_{65}}$ = 0.01 s$^{-1}$ |
| A38: $\frac{\partial Gene_{MRP_{ON}}}{\partial t}$ | $Gene_{MRP_{ON}(initial)}$ = 0 zmol | $k_{ind(NMA)_{65}}$ = -$^a$ | $k_{ind(XAAD)_{65}}$ = 1.32 zmol$^{-1}$.s$^{-1}$ |
| A39: $\frac{\partial Gene_{Nrf2_{OFF}}}{\partial t}$ | $Gene_{Nrf2_{OFF}(initial)}$ = 6.35E$^{-4}$ zmol | $k_{act_{27}}$ = 2.5E$^{-3}$ s$^{-1}$ | $k_{desact_{27}}$ = 0.01 s$^{-1}$ |
| A40: $\frac{\partial Gene_{Nrf2_{ON}}}{\partial t}$ | $Gene_{Nrf2_{ON}(initial)}$ = 0 zmol | $k_{ind(NMA)_{27}}$ = -$^a$ | $k_{ind(XAAD)_{27}}$ = 13 zmol$^{-1}$.s$^{-1}$ |
| A41: $\frac{\partial GS_{cytosol}}{\partial t}$ A42: $\frac{\partial GSmono_{cytosol}}{\partial t}$ | $k_{b_{38}}$ = 1.2E$^{-4}$ zmol$^{-1}$.s$^{-1}$ $GS_{cytosol(initial)}$ = 1632 zmol $GSmono_{cytosol(initial)}$ = 0 zmol | $k_{u_{38}}$ = 0.02 s$^{-1}$ $k_{TSL_{36}}$ = 0.0417 s$^{-1}$ | $k_{deg_{39}}$ = 1.93E$^{-5}$ s$^{-1}$ $k_{deg_{37}}$ = 3.86E$^{-5}$ s$^{-1}$ |

$^a$ : parameters estimated.

| Equation | Parameters and initial state variables | | |
|---|---|---|---|
| A43-A51: $\frac{\partial \gamma GC_{cytosol}}{\partial t}$ | $Gly_{cytosol}$ = 1.7E6 zmol | $K_{m1_{73}}$ = 11.2E5 zmol | $K_{m2_{73}}$ = 2.6E6 zmol |
| | $ATP_{cytosol}$ = 8.5E6 zmol | $Glu_{cytosol}$ = 1.7E7 zmol | $Cys_{cytosol}$ = 5.1E5 zmol |
| | $v_{max_{73}}$ = -[a] | $K_{m(Gly)_{73}}$ = 3E6 zmol | $K_{m(ATP)_{73}}$ = 12E4 zmol |
| | $K_{m(Glu)(GCL)_{72}}$ = 8.2E5 zmol | $v_{max(GCL)_{72}}$ = -[a] | $v_{max(GCLC)_{72}}$ = -[a] |
| | $\gamma GC_{cytosol_{(initial)}}$ = 2.6E5 zmol | $K_{m(ATP)(GCLC)_{72}}$ = 8.5E6 zmol | $K_{is(ATP)(GCLC)_{72}}$ = 2.2E6 zmol |
| | $K_{m(ATP)(GCL)_{72}}$ = 14.8E5 zmol | $K_{ii(ATP)(GCLC)_{72}}$ = 6.8E5 zmol | $K_{m(Glu)(GCLC)_{72}}$ = 2.7E6 zmol |
| | $K_{is(ATP)(GCL)_{72}}$ = 11E6 zmol | $K_{is(Glu)(GCLC)_{72}}$ = 5.1E5 zmol | $K_{ii(Glu)(GCLC)_{72}}$ = 14E5 zmol |
| | $K_{ii(ATP)(GCL)_{72}}$ = 6.6E6 zmol | $K_{m(Cys)(GCL)_{72}}$ = 3.7E5 zmol | $K_{m(Cys)(GCLC)_{72}}$ = 4.6E5 zmol |
| | $K_{is(Glu)(GCL)_{72}}$ = 14E5 zmol | $K_{ii(Glu)(GCL)_{72}}$ = 5.3E6 zmol | |
| A52: $\frac{\partial GSH_{cytosol}}{\partial t}$ | $GSH_{cytosol_{(initial)}}$ = 8.5E6 zmol | $v_{max_{74}}$ = 3137 zmol.s$^{-1}$ | $K_{m_{74}}$ = 3.4E7 zmol |
| | $K_{i(GSH)_{8b}}$ = 14.5E4 zmol | $K_{m(GSH)_{8b}}$ = 8.5E5 zmol | $v_{max_{8b}}$ = -[a] |
| | $K_{m(ROS)_{8b}}$ = 8.5E4 zmol | $K_{i(ROS)_{8b}}$ = 14.5E4 zmol | Refer to A51 for the other parameters. |
| A53: $\frac{\partial GST_{cytosol}}{\partial t}$ A54: $\frac{\partial GSTmono_{cytosol}}{\partial t}$ | $k_{b_{61}}$ = 3.4E-4 zmol.s$^{-1}$ | $k_{u_{61}}$ = 0.02 s$^{-1}$ | $k_{deg_{62}}$ = 1.29E-5 s$^{-1}$ |
| | $GST_{cytosol_{(initial)}}$ = 250 zmol | $k_{TSL_{59}}$ = 0.0417 s$^{-1}$ | $k_{deg_{60}}$ = 1.29E-4 s$^{-1}$ |
| | $GSTmono_{cytosol_{(initial)}}$ = 206 zmol | | |
| A55: $\frac{\partial Keap1_{cytosol}}{\partial t}$ A56: $\frac{\partial Keap1o_{cytosol}}{\partial t}$ | $Keap1_{cytosol_{(initial)}}$ = 34 zmol | $k_{red_{10}}$ = 0.1 s$^{-1}$ | $k_{ox_{10}}$ = -[a] |
| | $Keap1o_{cytosol_{(initial)}}$ = 0 zmol | $k_{b_{14}}$ = 3.4E-3 zmol.s$^{-1}$ | $k_{u_{14}}$ = 0.02 s$^{-1}$ |
| | $k_{u_{12}}$ = 0.014 s$^{-1}$ | $k_{b_{15}}$ = 3.4E-3 zmol.s$^{-1}$ | $k_{u_{15}}$ = 0.02 s$^{-1}$ |
| | $k_{u_{13}}$ = 1E-4 s$^{-1}$ | | |

[a] : parameters estimated.

| Equation | Parameters and initial state variables | | |
|---|---|---|---|
| A57: $\frac{\partial mRNA_{CYP}}{\partial t}$ | $mRNA_{CYP\,(initial)}$ = 0.787 zmol | $k_{TSP_{21}}$ = -[a] | $k_{deg_{22}}$ = 6E$^{-5}$ s$^{-1}$ |
| A58: $\frac{\partial mRNA_{GCLC}}{\partial t}$ | $mRNA_{GCLC\,(initial)}$ = 3.08 zmol | $k_{TSP_{42}}$ = -[a] | $k_{deg_{43}}$ = 4.83E$^{-5}$ s$^{-1}$ |
| A59: $\frac{\partial mRNA_{GCLM}}{\partial t}$ | $mRNA_{GCLM\,(initial)}$ = 1.33 zmol | $k_{TSP_{48}}$ = -[a] | $k_{deg_{49}}$ = 4.83E$^{-5}$ s$^{-1}$ |
| A60: $\frac{\partial mRNA_{GS}}{\partial t}$ | $mRNA_{GS\,(initial)}$ = 0.731 zmol | $k_{TSP_{34}}$ = -[a] | $k_{deg_{35}}$ = 4.83E$^{-5}$ s$^{-1}$ |
| A61: $\frac{\partial mRNA_{GST}}{\partial t}$ | $mRNA_{GST\,(initial)}$ = 0.731 zmol | $k_{TSP_{57}}$ = -[a] | $k_{deg_{58}}$ = 4.71E$^{-5}$ s$^{-1}$ |
| A62: $\frac{\partial mRNA_{GPx}}{\partial t}$ | $mRNA_{GPx\,(initial)}$ = 0.731 zmol | $k_{TSP_{57b}}$ = -[a] | $k_{deg_{58b}}$ = 4.71E$^{-5}$ s$^{-1}$ |
| A63: $\frac{\partial mRNA_{MRP}}{\partial t}$ | $mRNA_{MRP\,(initial)}$ = 6.46 zmol | $k_{TSP_{66}}$ = -[a] | $k_{deg_{67}}$ = 1.93E$^{-5}$ s$^{-1}$ |
| A64: $\frac{\partial mRNA_{Nrf2}}{\partial t}$ | $mRNA_{Nrf2\,(initial)}$ = 0.046 zmol | $k_{TSP_{28}}$ = -[a] | $k_{deg_{29}}$ = 6.43E$^{-5}$ s$^{-1}$ |
| A65: $\frac{\partial MRP_{cytosol}}{\partial t}$ A66: $\frac{\partial MRPmono_{cytosol}}{\partial t}$ | $k_{b_{70}}$ = 0.59E$^{-5}$ zmol$^{-1}$·s$^{-1}$ | $k_{u_{70}}$ = 0.02 s$^{-1}$ | $k_{deg_{71}}$ = 7.15E$^{-6}$ s$^{-1}$ |
| | $MRP_{cytosol\,(initial)}$ = 3.4E$^3$ zmol | $k_{TSL_{68}}$ = 0.0417 s$^{-1}$ | $k_{deg_{69}}$ = 1.93E$^{-5}$ s$^{-1}$ |
| | $MRPmono_{cytosol\,(initial)}$ = 3.4E$^3$ zmol | | |
| A67: $\frac{\partial MAF_{nucleus}}{\partial t}$ | $MAF_{nucleus\,(initial)}$ = 0.038 zmol | $k_{b_{18}}$ = -[a] | $k_{u_{18}}$ = 0.02 s$^{-1}$ |
| A68: $\frac{\partial NMA_{GCLC}}{\partial t}$ | $NMA_{GCLC\,(initial)}$ = 0 zmol | Refer to A3 for the parameters. | |
| A69: $\frac{\partial NMA_{GCLM}}{\partial t}$ | $NMA_{GCLM\,(initial)}$ = 0 zmol | Refer to A4 for the parameters. | |
| A70: $\frac{\partial NMA_{GS}}{\partial t}$ | $NMA_{GS\,(initial)}$ = 0 zmol | Refer to A5 for the parameters. | |
| A71: $\frac{\partial NMA_{GST}}{\partial t}$ | $NMA_{GST\,(initial)}$ = 0 zmol | Refer to A6 for the parameters. | |
| A72: $\frac{\partial NMA_{GPx}}{\partial t}$ | $NMA_{GPx\,(initial)}$ = 0 zmol | Refer to A7 for the parameters. | |
| A73: $\frac{\partial NMA_{MRP}}{\partial t}$ | $NMA_{MRP\,(initial)}$ = 0 zmol | Refer to A8 for the parameters. | |
| A74: $\frac{\partial NMA_{Nrf2}}{\partial t}$ | $NMA_{Nrf2\,(initial)}$ = 0 zmol | Refer to A9 for the parameters. | |

[a] : parameters estimated.

| Equation | Parameters and initial state variables |
|---|---|
| A75: $\frac{\partial Nrf2_{cytosol}}{\partial t}$ <br> A76: $\frac{\partial Nrf2_{nucleus}}{\partial t}$ | $Nrf2_{cytosol\,(initial)}$ = 0.85 zmol   $k_{deg_{31}}$ = 1E$^{-4}$ s$^{-1}$   $k_{TSL_{30}}$ = 0.0417 s$^{-1}$ <br> $Nrf2_{nucleus\,(initial)}$ = 0.038 zmol   $CL_{in_{16}}$ = 2 µm$^3$.s$^{-1}$   $CL_{out_{16}}$ = 1 µm$^3$.s$^{-1}$ <br> $k_{deg_{17}}$ = 1E$^{-4}$ s$^{-1}$    Refer to A55, A56, A64 and A67 for the other parameters. |
| A77: $\frac{\partial Nrf2\_Keap1_{cytosol}}{\partial t}$ <br> A78: $\frac{\partial Nrf2\_Keap1o_{cytosol}}{\partial t}$ | $Nrf2\_Keap1_{cytosol\,(initial)}$ = 0 zmol    $k_{red_{11}}$ = 0.1 s$^{-1}$ <br> $Nrf2\_Keap1o_{cytosol\,(initial)}$ = 0 zmol    Refer to A55, A56, A62 and A63 for the other parameters. |
| A79: $\frac{\partial Nrf2\_MAF_{nucleus}}{\partial t}$ | $Nrf2\_MAF_{nucleus\,(initial)}$ = 0 zmol    Refer to A3, A4, A5, A6, A7, A8 and A67 for the parameters. |
| A80: $\frac{\partial ROS_{cytosol}}{\partial t}$ | $ROS_{cytosol\,(initial)}$ = 0 zmol    $k_{f_{75}}$ = -$^a$    $k_{ROS}$ = -$^a$ <br> Refer to A52 for the other parameters. |
| A81: $\frac{\partial XAA_{nucleus}}{\partial t}$ | $XAA_{nucleus\,(initial)}$ = 0 zmol    Refer to A9, A17, A18, A19 and A20 for the parameters. |
| A82: $\frac{\partial XAAD_{CYP}}{\partial t}$ | $XAAD_{CYP\,(initial)}$ = 0 zmol    Refer to A18 for the parameters. |
| A83: $\frac{\partial XAAD_{GST}}{\partial t}$ | $XAAD_{GST\,(initial)}$ = 0 zmol    Refer to A19 for the parameters. |
| A84: $\frac{\partial XAAD_{MRP}}{\partial t}$ | $XAAD_{MRP\,(initial)}$ = 0 zmol    Refer to A20 for the parameters. |
| A85: $\frac{\partial XAAD_{Nrf2}}{\partial t}$ | $XAAD_{Nrf2\,(initial)}$ = 0 zmol    Refer to A21 for the parameters. |

$^a$ : parameters estimated.

Table S2: Cyclosporine A quantities measured in the extracellular medium (3mL) at low CsA concentration exposure (5 μM).

| Day | Time (in hr) | CsA quantity (in zeptomol) | | |
|---|---|---|---|---|
| | | Replicate 1 | Replicate 2 | Replicate 3 |
| 1 | 0.5 | 5.229E-06 | 5.105E-06 | 4.667E-06 |
| | 1 | 4.667E-06 | 5.000E-06 | 3.924E-06 |
| | 3 | 5.076E-06 | 4.695E-06 | 4.657E-06 |
| | 6 | 4.610E-06 | 4.743E-06 | 4.343E-06 |
| | 24 | 4.686E-06 | 4.352E-06 | 3.810E-06 |
| 3 | 24 | 5.781E-06 | 5.619E-06 | 5.210E-06 |
| 5 | 24 | 6.010E-06 | 5.771E-06 | 6.010E-06 |
| 7 | 24 | 5.076E-06 | 6.048E-06 | 5.771E-06 |
| 10 | 24 | 5.562E-06 | 5.410E-06 | 5.181E-06 |
| 14 | 0.5 | 4.905E-06 | 5.962E-06 | 5.219E-06 |
| | 1 | 5.333E-06 | 5.610E-06 | 5.076E-06 |
| | 3 | 5.124E-06 | 5.314E-06 | 4.419E-06 |
| | 6 | 5.943E-06 | 5.562E-06 | 5.162E-06 |
| | 24 | 5.238E-06 | 5.295E-06 | 4.971E-06 |

Table S3: Intracellular Cyclosporine A quantities measured. at low CsA concentration exposure (5 μM).

| Day | Time (in hr) | CsA quantity (in zeptomol) | | |
|---|---|---|---|---|
| | | Replicate 1 | Replicate 2 | Replicate 3 |
| 1 | 0.5 | 1.248E-06 | 1.181E-06 | 9.048E-07 |
| | 1 | 1.286E-06 | 1.152E-06 | 1.114E-06 |
| | 3 | 1.390E-06 | 1.362E-06 | 1.381E-06 |
| | 6 | 1.152E-06 | 1.248E-06 | 1.124E-06 |
| | 24 | 1.657E-06 | 1.305E-06 | 1.171E-06 |
| 14 | 0.5 | 1.771E-06 | 2.414E-06 | 2.086E-06 |
| | 1 | 3.271E-06 | 3.771E-06 | 3.886E-06 |
| | 3 | 2.600E-06 | 1.886E-06 | 2.100E-06 |
| | 6 | 1.914E-06 | 2.000E-06 | 2.543E-06 |
| | 24 | 2.043E-06 | 2.229E-06 | 2.514E-06 |

Table S4: Cyclosporine A quantities measured on plastic at low CsA concentration exposure (5 µM).

| Day | Time (in hr) | CsA quantity (in zeptomol) | | |
|---|---|---|---|---|
| | | Replicate 1 | Replicate 2 | Replicate 3 |
| 1 | 0.5 | 6.143E-07 | 7.095E-07 | 7.286E-07 |
| | 1 | 5.762E-07 | 6.429E-07 | 6.619E-07 |
| | 3 | 5.810E-07 | 7.667E-07 | 5.905E-07 |
| | 6 | 7.905E-07 | 8.333E-07 | 8.095E-07 |
| | 24 | 7.714E-07 | 1.024E-06 | 1.019E-06 |
| 14 | 0.5 | 1.010E-06 | 9.810E-07 | 8.952E-07 |
| | 1 | 9.048E-07 | 9.905E-07 | 1.062E-06 |
| | 3 | 1.414E-06 | 1.229E-06 | 1.462E-06 |
| | 6 | 9.238E-07 | 9.429E-07 | 8.762E-07 |
| | 24 | 7.524E-07 | 6.333E-07 | 8.000E-07 |

Table S5: Cyclosporine A quantities measured in the extracellular medium (3mL) at high CsA concentration exposure (15 µM).

| Day | Time (in hr) | CsA quantity (in zeptomol) | | |
|---|---|---|---|---|
| | | Replicate 1 | Replicate 2 | Replicate 3 |
| 1 | 0.5 | 1.888E-05 | 2.057E-05 | 1.693E-05 |
| | 1 | 1.686E-05 | 1.733E-05 | 1.636E-05 |
| | 3 | 1.557E-05 | 1.788E-05 | 1.674E-05 |
| | 6 | 1.555E-05 | 1.583E-05 | 1.745E-05 |
| | 24 | 1.581E-05 | 1.660E-05 | 1.490E-05 |
| 3 | 24 | 1.664E-05 | 1.355E-05 | 1.538E-05 |
| 5 | 24 | 1.674E-05 | 1.729E-05 | 1.605E-05 |
| 7 | 24 | 1.519E-05 | 1.612E-05 | 1.419E-05 |
| 10 | 24 | 1.652E-05 | 1.683E-05 | 1.757E-05 |
| 14 | 0.5 | 1.917E-05 | 2.076E-05 | 1.962E-05 |
| | 1 | 2.033E-05 | 2.007E-05 | 1.893E-05 |
| | 3 | 1.855E-05 | 1.886E-05 | 1.833E-05 |
| | 6 | 1.836E-05 | 1.850E-05 | 1.843E-05 |
| | 24 | 1.643E-05 | 1.612E-05 | 1.717E-05 |

Table S6: Intracellular Cyclosporine A quantities measured at high CsA concentration exposure (15 µM).

| Day | Time (in hr) | CsA quantity (in zeptomol) | | |
|---|---|---|---|---|
| | | Replicate 1 | Replicate 2 | Replicate 3 |
| 1 | 0.5 | 2.010E-06 | 1.671E-06 | 1.619E-07 |
| | 1 | 1.852E-06 | 2.086E-06 | 1.762E-06 |
| | 3 | 2.643E-06 | 2.500E-06 | 2.029E-06 |
| | 6 | 2.643E-06 | 2.310E-06 | 2.548E-06 |
| | 24 | 3.714E-05 | 2.548E-06 | 3.429E-06 |
| 14 | 0.5 | 2.222E-05 | 2.191E-05 | 1.696E-05 |
| | 1 | 2.160E-05 | 2.303E-05 | 2.371E-05 |
| | 3 | 1.350E-05 | 1.647E-05 | 1.548E-05 |
| | 6 | 1.603E-05 | 1.727E-05 | 2.315E-05 |
| | 24 | 1.467E-05 | 1.356E-05 | 1.418E-05 |

Table S7: Cyclosporine A quantities measured on plastic at high CsA concentration exposure (15 µM).

| Day | Time (in hr) | CsA quantity (in zeptomol) | | |
|---|---|---|---|---|
| | | Replicate 1 | Replicate 2 | Replicate 3 |
| 1 | 0.5 | 4.190E-07 | 5.238E-07 | 6.857E-07 |
| | 1 | 6.429E-07 | 8.782E-07 | 8.857E-07 |
| | 3 | 1.533E-07 | 9.000E-07 | 9.810E-07 |
| | 6 | 7.810E-07 | 1.143E-06 | 1.105E-06 |
| | 24 | 7.143E-07 | 1.090E-06 | 6.381E-07 |
| 14 | 0.5 | 5.905E-06 | 4.552E-06 | 5.000E-06 |
| | 1 | 6.390E-06 | 6.181E-06 | 7.067E-06 |
| | 3 | 8.752E-06 | 7.905E-06 | 4.848E-06 |
| | 6 | 6.248E-06 | 4.829E-06 | 4.448E-06 |
| | 24 | 5.086E-06 | 6.305E-06 | 5.762E-06 |

Table S8: Fold changes measured at low CsA concentration (5 µM).

| Species | Day 1 | | | Day 3 | | | Day 14 | | |
|---|---|---|---|---|---|---|---|---|---|
| | R1 | R2 | R3 | R1 | R2 | R3 | R1 | R2 | R3 |
| CYP mRNA | 1.115 | / | / | 1.050 | / | / | 0.945 | / | / |
| GCLC mRNA | 1.043 | 1.046 | / | 1.093 | 1.038 | / | 1.022 | 0.995 | / |
| GCLM mRNA | 0.999 | 0.897 | / | 1.198 | 1.169 | / | 1.168 | 1.316 | / |
| GPx mRNA | 0.909 | / | / | 0.930 | / | / | 0.685 | / | / |
| GS mRNA | 0.951 | / | / | 0.886 | / | / | 0.971 | / | / |
| GST mRNA | 1.069 | 1.054 | / | 0.948 | 0.981 | / | 1.081 | 1.033 | / |
| MRP mRNA | 1.086 | / | / | 1.093 | / | / | 1.133 | / | / |
| Nrf2 mRNA | 0.935 | / | / | 1.329 | / | / | 1.156 | / | / |
| GCLM | / | / | / | 1.069 | / | / | / | / | / |
| GS | 0.949 | 1.010 | 1.020 | 0.952 | 0.964 | 1.017 | 0.971 | 1.003 | 1.040 |
| GSH | 1.092 | 1.702 | 0.701 | 1.416 | 0.799 | 1.428 | 1.945 | 0.932 | 1.468 |
| γ-GC | 4.602 | 39.391 | 2.869 | 5.192 | 1.538 | 2.135 | 0.542 | 2.749 | 0.232 |
| MRP | / | / | / | / | / | / | 0.782 | / | / |

Table S9: Fold changes measured at high CsA concentration (15 µM).

| Species | Day 1 | | | Day 3 | | | Day 14 | | |
|---|---|---|---|---|---|---|---|---|---|
| | R1 | R2 | R3 | R1 | R2 | R3 | R1 | R2 | R3 |
| CYP mRNA | 1.099 | / | / | 1.004 | / | / | 1.211 | / | / |
| GCLC mRNA | 3.425 | 1.299 | / | 1.945 | 1.211 | / | 2.478 | 1.206 | / |
| GCLM mRNA | 2.755 | 4.709 | / | 2.848 | 5.154 | / | 1.769 | 2.785 | / |
| GS mRNA | 0.764 | / | / | 0.728 | / | / | 0.816 | / | / |
| GPx mRNA | 4.622 | / | / | 2.478 | / | / | 1.476 | / | / |
| GST mRNA | 1.056 | 1.077 | / | 1.010 | 0.980 | / | 0.986 | 0.974 | / |
| MRP mRNA | 1.317 | / | / | 1.194 | / | / | 1.081 | / | / |
| Nrf2 mRNA | 1.133 | / | / | 1.408 | / | / | 1.261 | / | / |
| GCLM | / | / | / | 1.431 | / | / | / | / | / |
| GS | 0.971 | 1.035 | 0.973 | 0.955 | 0.988 | 1.040 | 0.943 | 0.980 | 1.007 |
| GSH | 2.093 | 1.423 | 4.528 | 7.063 | 8.376 | 4.216 | 5.017 | 2.899 | 6.975 |
| γ-GC | 30.13 | 181.4 | 11.38 | 92.085 | 525.6 | 32.53 | 14.237 | 39.93 | 6.953 |
| MRP | / | / | / | / | / | / | 0.775 | / | / |

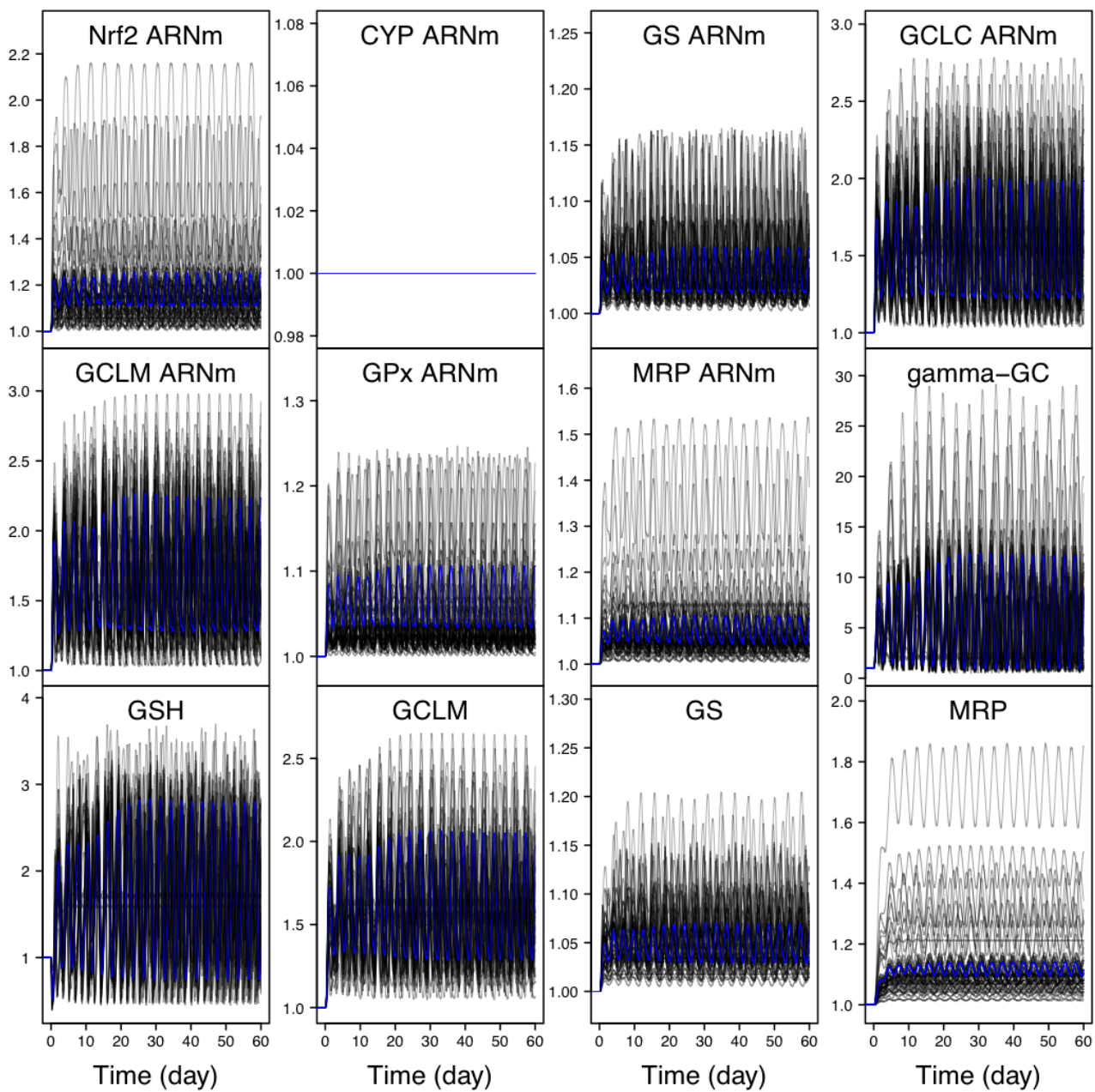

**Figure S1**: Transcriptomics (Nrf2 mRNA, CYP mRNA, GS mRNA, GCLC mRNA, GCLM mRNA, GPx mRNA, and MRP mRNA) proteomics (GCLM, GS, and MRP), and metabolomics (γ-GC, and GSH) fold-changes time-course in RPTEC cells during 60 days with repeated low dose (5 μM) CsA dosing. The blue line indicates the best fitting (maximum posterior probability) model prediction. The black lines are predictions made with 49 random parameter sets.

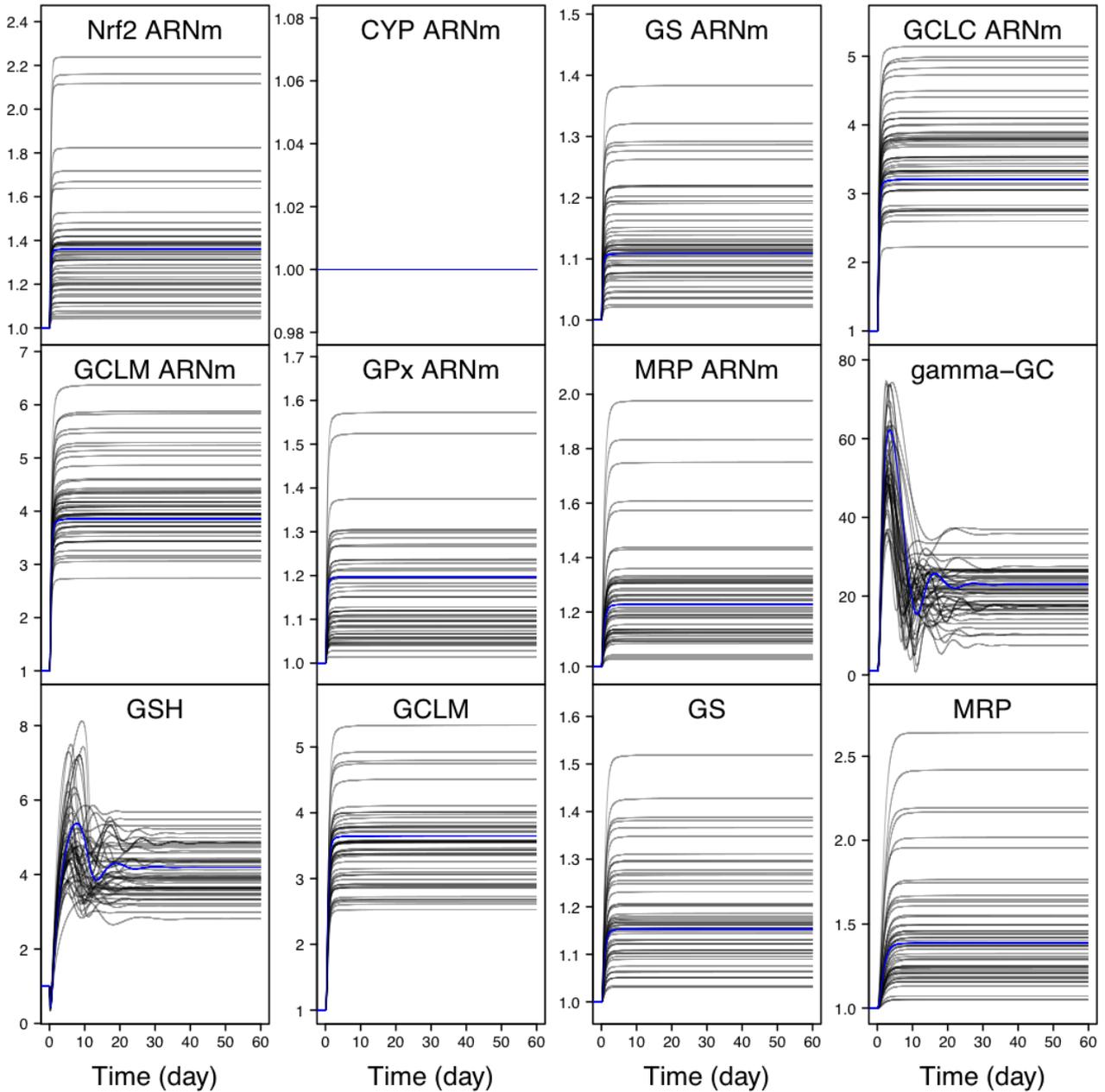

**Figure S2**: Transcriptomics (Nrf2 mRNA, CYP mRNA, GS mRNA, GCLC mRNA, GCLM mRNA, GPx mRNA, and MRP mRNA) proteomics (GCLM, GS, and MRP), and metabolomics (γ-GC, and GSH) fold-changes time-course in RPTEC cells during 60 days with repeated high dose (15 μM) CsA dosing. The blue line indicates the best fitting (maximum posterior probability) model prediction. The black lines are predictions made with 49 random parameter sets.